# A non-contact mutual inductance based measurement of an inhomogeneous topological insulating state in Bi$_2$Se$_3$ single crystals with defects

Amit Jash[1], Kamalika Nath[1], T. R. Devidas[2,3], A. Bharathi[2], S. S. Banerjee[1*]

[1]Department of Physics, Indian Institute of Technology, Kanpur 208016, Uttar Pradesh, India; [2]UGC-DAE Consortium for Scientific Research, Kalpakkam-603104, India; [3]Present address; The Racah Institute of Physics, Hebrew University of Jerusalem, Givat Ram, Jerusalem, 91904, Israel

**Abstract:** Pure Topological Insulating materials preserve a unique electronic state comprising of bulk insulating gap and conducting surface states. Here we use bulk Bi$_2$Se$_3$ single crystals possessing Se vacancy defects as a prototype topological insulator (TI) material for exploring the effect of non-magnetic disorder on the conducting properties of TI. We employ a sensitive, non-contact mutual inductance based technique for measuring the surface and bulk contribution to electrical conductivity in the TI. We discern the bulk and surface contributions by observing that predominant surface electrical conduction shows linear frequency dependence of the pickup signal while bulk conductivity gives rise to quadratic frequency dependence. We also observe an algebraic temperature dependent surface conductivity and an activated form of bulk electrical conductivity. Using the above we uncover an interplay between surface and bulk contribution to electrical conductivity in the TI as a function of temperature. In the Bi$_2$Se$_3$ crystals the transformation from surface to bulk dominated electrical transport is found to occur close to a temperature of 70 K. This temperature matches well with our results from activated bulk electrical transport results which shows an activation energy scale Δ which is in the meV range. The gap Δ is much less than the bulk band gap in Bi$_2$Se$_3$ which we argue is associated with defect states in the TI material. To understand our results, we propose a model of TI comprising of an inhomogeneous low electrically conducting medium (bulk) which is sandwiched between thin two high electrically conducting sheets (surface). The inhomogeneous TI state we argue is generated

---

by Selenium vacancies defects in $Bi_2Se_3$, which is responsible for producing an interplay between bulk and surface conductivity.

## I. Introduction:

In recent times, new materials like three dimensional (3D) topological insulator, exhibit topologically protected bulk gapped state enclosed by conducting surface state [1,2,3,4,5,6]. Herein no continuous variation of the lattice parameter can change the unique character of the electronic states in these materials. These gapless surface states in a topological insulator (TI) exhibit Dirac-like linear energy - momentum dispersion [7,8], chiral spin texture [5] and Landau level quantization [9]. Materials like $Bi_2Se_3$, $Bi_2Te_3$ and $Bi_{1-x}Sb_x$ are amongst the most well studied three dimensional TIs [2,7,10]. The characteristic TI properties have been confirmed via different techniques like angle-resolved photoemission spectroscopy (ARPES) [3,5,11,12], Shubnikov-de Haas (SdH) oscillations, [13,14] scanning tunneling microscopy (STM) and transport studies [2,15]. Spin momentum locking in TI causes chiral currents to flow which are unaffected by scattering from disorder [1,2,15,16]. Recent development of pump probe techniques to study spin momentum locking [17] reveal that TI are important from the point of view of spintronics and quantum computation applications [18,19,20]. While pristine TI are important, it is also important to explore how the TI state is affected in the presence of material defects. While presence of magnetic impurities will lead to breaking time reversal symmetry and hence destroy the TI state, the issue of how non-magnetic disorder affects the TI state is worth investigating.

To study the effect of disorder on TI, the popular $Bi_2Se_3$ crystals are a suitable choice, as they intrinsically possess Se vacancies which are a nature source of disorder in this material. The TI nature of $Bi_2Se_3$ has been identified through magneto-transport studies which confirm the presence of conducting surface states through the observation of SdH oscillations and weak anti-localization effects [16,21,22,23,24,25]. In $Bi_2Se_3$ the bulk insulating gap has been estimated to be ~ 300 meV [7,10,26,27]. However the presence of Se vacancies electron dopes the material [28,29,30], thereby enhancing the bulk electrical conductivity. The resulting parallel conducting channels through the surface and bulk, make it difficult to explore the details of the individual bulk and surface state contributions to electrical conductivity in these TI materials [6,31,32]. In such materials, the surface states have been identified either by suppressing defect induced bulk conductivity through counter doping [7,15] or by comparing changes in the conducting properties as a function of the thickness of TI thin films [21,22,23,33,34]. Without counter doping, the issue of investigating how Se vacancies affects the TI state in $Bi_2Se_3$ single crystals is none the less important. Studies show that increasing Se vacancy concentration weakens the SdH oscillations, which is a fingerprint of the conducting surface states characterizing a TI [21-24,33,34,35,36]. The SdH oscillations in magneto-transport are typically measurable at low temperature and high magnetic field regime. Hence, due to the limitations of the transport studies mentioned above, while the presence of disorder like Se vacancies seem to weaken the TI state, the details of how exactly the surface and bulk conductivities are affected by disorder and modified by Se vacancies, still remains unclear. Furthermore, studies have shown that due to degradation of surface of $Bi_2Se_3$, for example due to oxidation of the surfaces, causes band bending of the surface states extending upto 20 nm into the bulk [37,38]. Hence, there is a need for a technique which distinguishes between bulk and surface contributions to electrical conductivity in a TI. We report here a non-contact measurement technique of shielding currents induced in a TI using a two coil mutual inductance setup. We study the frequency ($f$) and temperature ($T$) dependence of the pickup voltage, from five single crystals of different thickness of $Bi_2Se_3$ placed between the coils. The pickup voltage shows two distinct regimes of frequency dependence, viz., a quadratic and linear regime. We show that the quadratic frequency dependence is associated with bulk electrical conductivity in the TI while the linear frequency dependence regime is associated with surface conductivity in the TI. As a function of temperature, we show that below 70 K, the frequency dependence of the pickup voltage is predominantly linear while it turns quadratic at higher $T$. Analysis of the data also shows that below 70 K we identify an algebraic temperature dependent electrical conductivity regime which saturates at low temperatures, while above 70 K electrical conductivity is of thermally activated type with a thermal activation energy scale ($\Delta$) of tens of meV. From our study we identify four distinct temperature regimes for the $Bi_2Se_3$ crystal, identifying the crossover between different surface and bulk dominated conductivity regimes. At high temperature above 180 K, we observe the unusual return of surface dominated conductivity ($\sigma_s$) coexisting with bulk conductivity ($\sigma_b$) in the material. To understand our results, we propose a simplified model of a TI, comprising of a low conducting bulk sandwiched between two high conducting surface sheets. Our simulations show surface and bulk conductivity values in the $Bi_2Se_3$ crystals to be of the order of $10^{11}$ S/m and $10^3$ S/m, respectively. While this minimal model explains the overall features of the data and offers a way to estimate the surface and bulk electrical conductivities of the TI from the measurements, it doesn't match the data exactly. A better agreement with the data is obtained by considering an inhomogeneous TI state. Inhomogeneous TI is modelled by incorporating in the minimal model, conducting channels threading the low conducting bulk medium. The model fits our experimental data by introducing inhomogeneity 30 percent. The inhomogeneous TI state we argue is a result of disorder in the TI bulk, generated by Selenium vacancies. Excess charge carriers produced by Se vacancies produce the gap ~ $\Delta$, which results in an interplay between bulk and surface conductivity.

## II. Experimental details: Transport measurement of $Bi_2Se_3$ single crystals.

For our study we use single crystals of $Bi_2Se_3$ prepared by slow cooling stoichiometric melts of high purity bismuth (Bi) and selenium (Se) powders [for details see refs.35,36]. In this work we have investigated the surface and bulk conductivities across five different single crystals belonging to the same batch of $Bi_2Se_3$ crystals grown (with similar electrical transport characteristics, for example see Fig.1). The thickness (surface area) of the crystals are, 20 μm (2.4 mm × 1.9 mm) (hence forth referred to as S20 sample), 51 μm (S51) (2.8 mm × 2.5 mm), 69 μm (S69) (3.9 mm × 2.5 mm), 75 μm (S75) (3.7 mm × 2.6 mm) and 82 μm (S82) (3.2mm × 2.8 mm). The electrical transport measurements shown in Fig.1 are on a $Bi_2Se_3$ single crystal with thickness of 70 μm. Note that as transport measurements involves making of electrical contacts on the sample which introduce irreversible physical changes in the sample, hence for our non contact measurements (which is the main topic of our manuscript) and transport measurements (here and refs.[35,36]), we are compelled to use different samples for the two measurements, albeit chosen from the same batch of single crystals grown with similar sample thickness. Figure 1(a) shows distinct SdH oscillation in longitudinal magneto-resistance ($R_{xx}$ vs magnetic field ($B$)) measurements at different temperature using standard Van der Pauw geometry. The low temperature SdH oscillations which weaken with increasing $T$, indicates the topological character of conducting surface states in TI at low $T$ [13,14,21,22,23,35,36] (more detailed analysis of the transport data is shown elsewhere [35,36]). The figure shows that the topological character of transport progressively gets affected with increasing $T$. The Berry phase ($\phi$) is calculated from the LK fitting and the value being close to $\pi$ at low $T$, suggests the SdH oscillation arises from surface Dirac electron (see inset of Fig. 1(b), and discussed in later section). The destruction of the SdH oscillations is identified with the onset of bulk contribution to electrical conductivity. Figure. 1(b) shows the Hall resistance ($R_{xy}$) measurement as a function of magnetic field ($B$) at different temperatures. This measurement establishes electrons as the charge carriers in these $Bi_2Se_3$ samples. It may be recalled that electrical transport measurement studies have shown that Se vacancies lead to weakening the SdH oscillations [35,36]. Recent studies show a correlation between Positron annihilation lifetime with Se vacancy concentration in $Bi_2Se_3$. These studies, performed on this batch of samples show Se vacancy concentration are in the range of $10^{17}$ cm$^{-3}$ [35]. From these SdH measurements at low $T$ and high $B$ it is however difficult to determine the extent of surface and bulk contributions to conductivity, and hence study how it is affected by either $T$ or disorder variation present either in on the surface or bulk of the material (note, we shall return to this issue in the discussion section). As mentioned earlier the appearance of finite bulk conductivity in $Bi_2Se_3$ sample with Se vacancies mixes the surface and bulk contributions to electrical conductivity making it difficult to discern the individual contributions. In the next section, we discuss a non-contact measurement technique which allows us to distinguish between the two contributions to electrical conductivity.

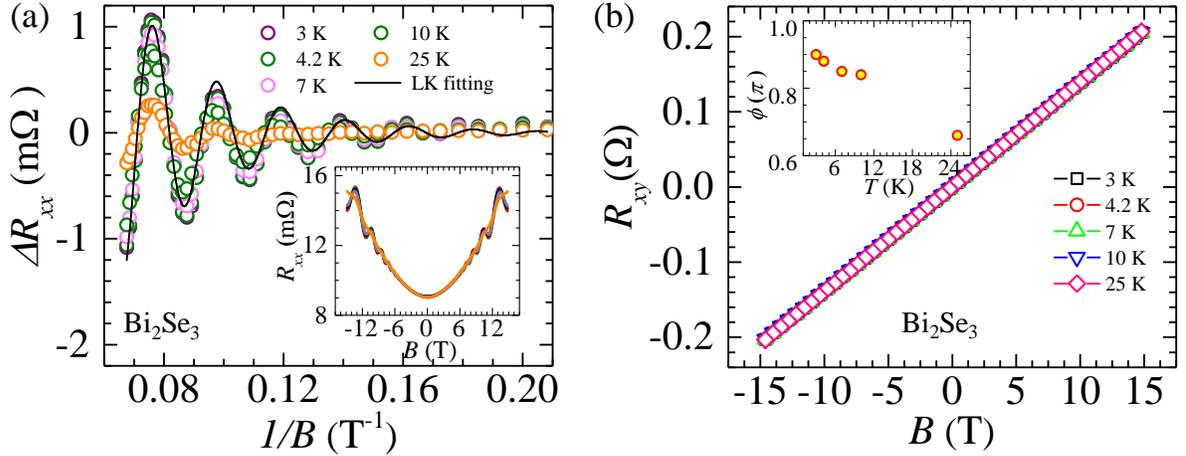

FIG. 1. (a) Inset shows the variation of $R_{xx}$ as a function of magnetic field $B$ measured in the standard Van der Pauw geometry at 3 K, 4.2 K, 7 K, 10 K and 25 K (see legends in the main panel). The main panel shows $\Delta R_{xx}$ vs. $1/B$ plot showing SdH oscillations in $Bi_2Se_3$ (sample thickness 70 μm) at different temperatures. $\Delta R_{xx}$ is calculated by subtracting from the experiment $R_{xx}(B)$ values a polynomial fit to the data ($R_{poly}(B)$), where $B$ is the magnetic field. The data is fitted to, $R_{poly}(B) = R_0 + R_1 B + R_2 B^2$, where $R_0$=9.26x10$^{-3}$ Ω, $R_1$=-1.71x10$^{-6}$ Ω.T$^{-1}$, $R_2$=2.91x10$^{-5}$ Ω.T$^{-2}$. Lifshitz-Kosevich (LK) equation is used to fit the $\Delta R_{xx}$ data at 4.2 K which is shown by black solid line. For fitting to the magneto-resistance data in Fig. 1(a), following ref. [35] and the values therein, we use

$$\Delta R_{xx} = a\sqrt{0.011B}\left(\frac{\frac{11.12}{B}}{\sinh\left(\frac{11.12}{B}\right)}\right)e^{-\frac{19.38}{B}}(0.95)\cos\left[2\pi\left(\frac{F}{B}+\beta\right)\right].$$ In this equation $a = 0.0135$ Ω and $F$ and $\beta$ are the fitting parameters. A fit to the data (see black solid line in Fig. 1(a)), gives $F= 46.95 \pm 0.25$ T and $\beta = 0.43$. (b) Hall measurement of the transverse resistance, $R_{xy}$ vs $B$ at different temperatures. The sample used in the above measurements has thickness ~ 70 μm. Inset shows the Berry phase as function of temperature which is measured from the phase of the LK equation, viz., $\phi = 2\pi\beta$.

### III. Two coil mutual inductance measurement setup: Non-contact measurement

To characterize the surface and bulk states of TI we use a modified non-contact two coil mutual inductance measurement technique. The two coil mutual inductance technique is well-established experimentally [39,40,41,42] and it is sensitive enough to experimentally study the shielding response of superconductors, especially in situations when it becomes weak like that near the critical temperature of a superconductor. The technique has been analyzed theoretically to determine the nature of the field and current distributions generated in the sample placed within the two coil configuration, by incorporating effects of skin depth or penetration depth [43,44,45] and studying the behavior of the pickup signal generated. Note that, whereas in conventional electrical transport measurement the Joule heating at electrical contacts complicate temperature dependent conductivity measurements, here such issues are avoided through our non-contact two-coil mutual inductance measurement technique. Figure 2(a) shows the schematic of our setup where a crystal is placed between an excitation coil and a pickup coil. Both coils are designed with very closely matched coil parameters (for details on coil dimensions and other parameters see S1 of the supplementary information in Ref. [46] ). The data for all our pickup measurements corresponds to AC current of amplitude 153 mA applied to the excitation coil (corresponding to 1Volt drop across the coil) at frequency $f$, and the real and imaginary components of the pickup voltage from the pickup coil is measured using a lock-in amplifier. Measurements at higher AC excitation current amplitude, yield similar results. In the supplementary section S1 (Ref. [46]) we show the linear relationship between the excitation current in the coil and the pickup voltage for an S20 sample at 65 kHz excitation frequency. The excitation current generates a time-varying magnetic field, which induces currents inside the conducting sample, that in-turn leads to a time-varying magnetic field associated with the sample. This local field induces a voltage in the pickup coil, which is being measured. Effectively one considers that the presence of a sample between the two coils modifies the mutual inductance. Variations in the sample properties change the signal induced in the pickup coil. One may note that voltage is induced in the pickup coil not only from time-varying magnetic fields associated with the sample, but also the via the stray magnetic fields present outside the sample. These stray field induced pickup voltage signal can often lead to a large background signal which masks the signal from the sample if it is weak. In order to significantly reduce this stray field effect between the dipolar coil assembly, a 1.5 mm thick oxygen-free high thermal conductivity Cu (OFHC) sheet (resistivity ~ $1.7 \times 10^{-8}$ $\Omega$.m) with a hole at its center (see Fig. 2(a)), is placed coaxially above the excitation coil. The hole diameter (2 mm) is chosen such that it doesn't exceed the sample surface dimensions (the sample is placed above the Cu sheet covering the hole). Note that the OFHC Cu sheet thickness (1.5 mm) is larger than its skin depth, which at 60 kHz is 0.27 mm and at 5 kHz is 0.92 mm. The high conductivity thick OFHC Cu sheet shields the alternating magnetic field generated by the excitation coil, except over the hole. The electromagnetic shielding provided by the thick Cu sheet helps to reduce the stray flux linkage outside the sample, while the hole in the sheet concentrates the magnetic flux on the sample. Therefore, the Cu sheet with the hole effectively enhances the coupling of the two coils via the sample in between and the effects of the stray field is minimized. In Figs. 2(b) and 2(c), the colored regions represent the simulated vertical ($z$) component of the AC magnetic field distribution ($B_z$) around excitation coil (for simulation details, see S2 section in supplementary information of Ref. [46]). Figure 2(c) shows the concentration of magnetic flux above the hole in the Cu plate. In Fig. 2(d), the simulated $B_z$ profile measured above the coil (black dashed line in Fig. 2(c)), shows the significant concentration of magnetic flux above the hole in the Cu sheet. The sample when placed above the Cu sheet experiences this concentrated oscillating magnetic field and the oscillating magnetic field induces a voltage in the pickup coil voltage. Figure 2(c) shows that due to the hole in Copper sheet, the central field (amplitude of the AC field) within the coil above the hole is increased by almost a factor of two compared to the field above the solid Cu sheet outside the hole. We thus achieved almost a two times increase in the signal induced within the pickup coil due to the use of the thick copper sheet with hole. When the sample which is slightly larger than the hole diameter is placed over the hole in the Cu sheet, then due to the enhancement in the field over the hole, a stronger signal emanates preferentially from the sample which induces a voltage in the pickup coil. Thus by using the hole in the Cu sheet we obtain a signal which is preferentially emanating from the sample. Without the use of a Cu sheet with hole, viz., without any Cu sheet the Fig. 2(d) (see green curve) shows the field is uniformly strong over within the entire coil area. In this configuration if a sample is placed between the coil and the magnetic susceptibility of the sample isn't high, then the signal induced in the coil would get buried in the background signal. Here the background signal emanates from the time varying magnetic flux in regions outside the sample area which induce voltage in the pickup coil apart from the signal induced by the sample. Hence the use of the Cu sheet with the hole, preferential by concentrates the magnetic flux over the sample area, which therefore helps to induce a stronger signal in the pickup coil emanating from the sample. The use of the Cu sheet also helps to significantly reduce the stray field coupling from outside the coil area as the thick Cu sheet electromagnetically shields the alternating AC magnetic field and thereby reduces the background signal outside the sample further. This modification enhances the sensitivity of the two coil configuration in measuring the magnetic response of the sample compared to the conventional configuration without the Copper sheet with hole. During the measurements with TI we balance out the imaginary part of the signal as it doesn't change significantly during measurements. Furthermore, we would like to emphasize that for all our measurements we subtract the background voltage at all temperature and frequencies by measuring the pickup signal without a sample and with only the Cu sheet with a hole placed between the coils.

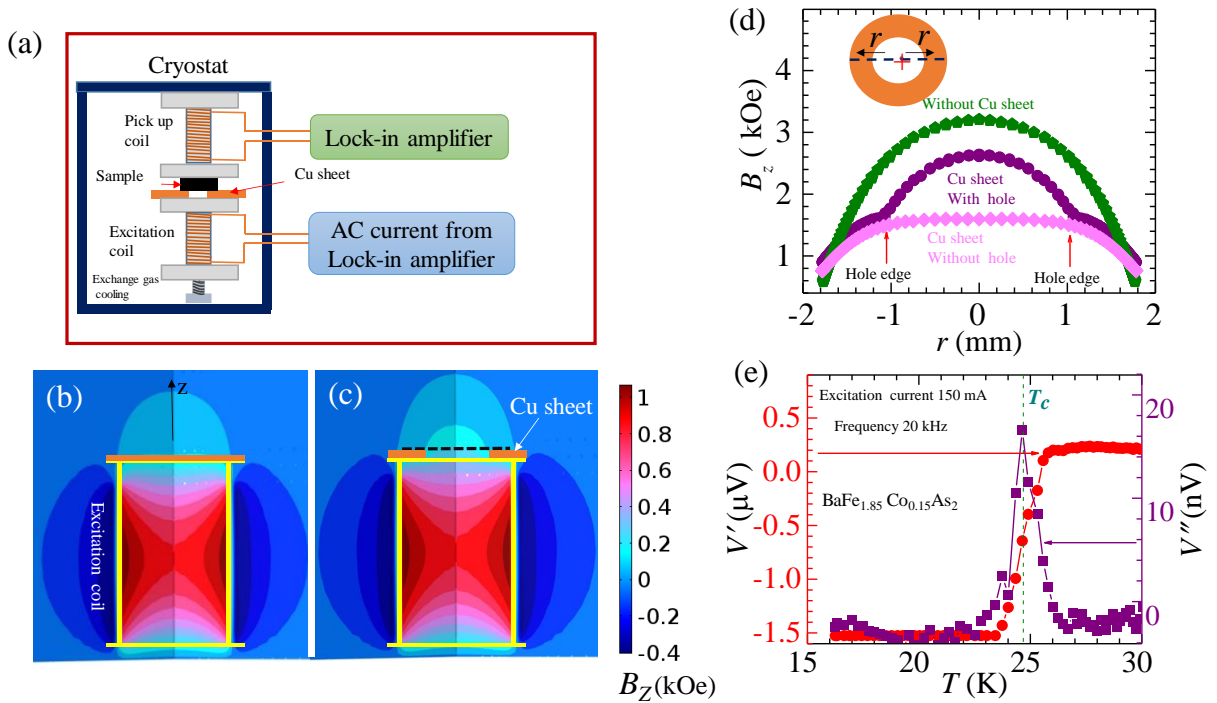

FIG. 2. (a) Schematic diagram of the two coil mutual inductance setup is shown. (b) & (c) Show the simulated normal component of the AC magnetic field ($B_Z$) distribution for the two cases. The outline of the solid excitation coil bobbin is shown in yellow. In Fig. (b) only a solid Cu sheet (orange) is placed above the excitation coil and in Fig. (c) a Cu Sheet with a hole at the center is placed over the coil. The sample is placed above the hole in the Cu sheet. These simulations are performed using COMSOL with 150 mA, AC current at 60 kHz sent through the excitation coil. (d) Shows the simulated $B_z$ vs $r$ profile as measured across the black dashed line shown in Fig. (c) (r = 0 is the central (z) axis of the excitation coil bobbin), for three different cases: (i) without any Cu sheet above the pickup coil, (ii) Cu sheet with a hole, (iii) Cu sheet with no hole. The inset of Fig. (d) shows the schematic of the circular Cu sheet with a hole placed over the top of the excitation coil. (e) The superconducting transition of Iron Pnictide single crystal, $BaFe_{1.85}Co_{0.15}As_2$ is shown. The figure depicts the behavior of the real (left axis) and imaginary (right axis) component of the pickup voltage as a function of temperature. The $BaFe_{1.85}Co_{0.15}As_2$ crystal dimension is 3.2 mm × 2.4 mm × 0.5 mm. The $T_c$ is indicated as the green dashed line ~ 24.7± 0.2 K. A similar $T_c$ value of this sample is observed through bulk magnetization measurement on a SQUID magnetometer.

We test the performance of the setup using a superconducting sample as its shielding response is well known. Figure 2(e) illustrates the performance of our setup by measuring the AC susceptibility response of a superconducting single crystal, viz., an optimally doped Iron Pnictide crystal ($BaFe_{1.85}Co_{0.15}As_2$) which is placed above the hole in the Cu sheet. Due to strong superconducting diamagnetic shielding of AC magnetic field, (Fig. 2(e)) we see the rapid drop in the in phase signal in the pickup coil ($V'$) at $T < T_c$. Near $T_c$, we also see the expected peak in the out of phase signal ($V''$). To identify the negative signal below transition temperature, we have subtracted the data form the saturation value above $T_c$. The superconducting transition temperature $T_c$ estimated from the peak position in $V''$ is 24.7 K which compares well with $T_c$ reported for this stoichiometry [47,48]. The observation in Fig. 2(e) that at the characteristic temperature $T_c$, there is a sharp change in the background corrected, $V'$ signal (viz., the diamagnetic shielding response) and $V''$ signal (viz., the dissipation response), suggests that these feature cannot be related to features associated with the background signal (when no sample is present and only the Cu sheet with hole is present). The behavior of the background signal of the two coil setup (with only the Cu sheet placed between the hole) is simulated and pickup voltage signal due to solid Cu sheet is shown in the supplementary information (see S3 section in supplementary information of Ref. [46]). The close match between the experimentally measured background signal with the simulated data (S3 section in supplementary information of Ref. [46]) shows that the background signal can be explained using standard Maxwell's equations for a metallic conductor. There is nothing in the equations which can cause a non-monotonic change in the background signal.

## IV. Studying the shielding response of Bi$_2$Se$_3$ single crystal at different $f$ and $T$:

Figure 3(a) shows the $V(f)$ measured for the S69 Bi$_2$Se$_3$ sample at different temperature, where $V(f) = \sqrt{V'^2(f) + V''^2(f)} \approx V'(f)$, as for these TI crystals, $V''(f) \ll V'(f)$. In the figure One can identify two regimes of behavior in $V(f)$, namely, one in the low frequency regime where $V(f) \propto f^2$ (see red dotted line through the data), and the other in the higher frequency regime where, $V(f) \propto f^\alpha$ (where $\alpha = 0.9 \pm 0.05$) (see black solid line to the data). The $V(T)$ data in Fig. 3(b) shows that the pickup voltage saturates to constant value at low $T$ upto 40 K. Above 40 K to 70 K the $V(T)$ data fits (solid black line) with $\sigma_s(T) = 1/(C+DT)$ where $C$ is related to static disorder scattering and $D$ to electron-phonon coupling strength. While above 70 K to 170 K the data fits (red dashed line) with $\sigma_b(T) = \sigma_{b0} \exp\left(-\Delta/K_bT\right)$, $\Delta \sim 25.2 \pm 1.25$ meV, where $\Delta$ is the activation energy scale and $\sigma_{b0}$ is the high temperature conductance of the bulk state. Above the temperature 170 K, the data deviates from the thermally activated behavior which is present between 70 K to 170 K. Figure 3(c) shows that $V(f)$ is linear over a wide frequency regime at $T \leq$ 40 K. Inset of Fig. 3(a) shows that at 40 K, $V(f)$ increases linearly with $f$ for the entire frequency range. At 220 K (see Fig. 3(a) inset) $V(f)$ is not quadratic rather $V(f) \propto f^\alpha$ above 20 kHz, where $\alpha$ ~1, viz., almost linearly dependent. Between 70 K & 170 K $V(f) \propto f^2$ for $f \leq 20$ kHz and $V(f) \propto f^\alpha$ ,where $\alpha$ ~1 for higher frequency. Identical features at different $f$ and $T$ are shown in supplementary for other samples as well (see supplementary sections S4, S5, S6, S7 & S8 of Ref. [46]).

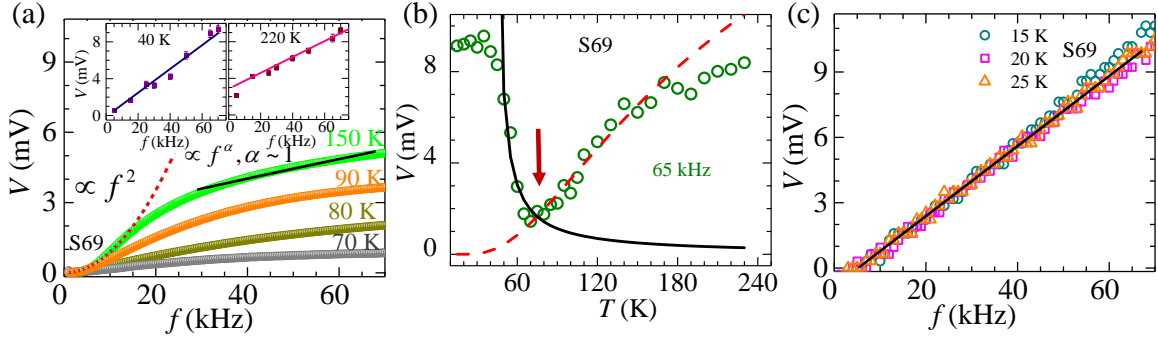

FIG. 3. (a) Shows *V(f)* responses at higher temperatures, 150 K, 90 K, 80 K, 70 K (from top to bottom curves). The quadratic ($V(f) \propto f^2$) and nearly linear ($V(f) \propto f^\alpha$, where $\alpha = 0.9 \pm 0.05$) regions are shown by the dotted red line and solid black line respectively. Insets show pickup response at 40 K and 220 K (*V* is measured from *V(T)* scan at different frequencies). For both 40 K and 220 K above 20 kHz, *V* is increasing linearly with frequency. All data in the figures correspond to excitation current of 153 mA in the excitation coil. (b) Variation of pickup voltage (*V*) with temperature (*T*) for sample S69 at frequency 65 kHz. Red dotted line is the fitted line corresponding to bulk conductivity, viz., $V \propto V_{b0}\, exp\left(-\Delta/K_b T\right)$ and the black line is fitted with surface conductivity, i.e., $V \propto 1/(C' + D'T)$. The fitting parameters are $C' = 0.0052$ mV$^{-1}$, $D' = 1.86$ mV$^{-1}$K$^{-1}$, $V_{b0} = 18.7$ mV and $\Delta$ is $(25.2 \pm 1.25)$ meV. (c) The behavior of pickup voltage with frequency is plotted at different low temperatures (15 K, 20 K, 25 K).

We would like to mention that the pickup signal, at low *f* is lesser than that at 70 kHz in a frequency dependent measurement or it is lesser at low *T* compared to that above 300 K in a *T* dependent measurement. Therefore, while measuring the pickup signal as a function of *f* or *T* one can balance out the signal above, 70 kHz or 300 K, to note the negative sign of the pickup signal corresponding to the diamagnetic nature of $Bi_2Se_3$.

## V. Understanding the frequency dependence of the pickup signal:

To understand the above frequency dependence of the pickup voltage, we recall that AC magnetic field produced from the excitation current in the primary induces screening currents extending upto different depths inside the conducting TI sample (skin depth). The magnetic field generated from the sample by these induced screening currents couple with the pickup coils to induce a pickup voltage. It is known that the depth upto which the currents are induced in the sample depends on the frequency of AC field. At low frequencies as the AC excitation penetrates deeper into the bulk of the sample (due to large skin depth), hence the bulk properties in the TI sample are probed with low *f*. The schematic of the distribution of the magnitude of induced screening current ($I_{induced}$) across the sample cross section is shown in Figs 4(a) and 4(b) for low frequency and high frequency respectively. The expression for skin depth ($\delta$) is $\sqrt{\dfrac{1}{\pi \sigma f \mu}}$, where $\mu$ is permeability of $Bi_2Se_3$. At high frequency Fig. 4(b), screening currents circulate within the high conductivity surfaces. We show in S11 in supplementary section, Ref. [46] a simulation showing the attenuation of an impinging EM signal is governed by the high conducting surface sheath in the TI material. Hence the properties of the surfaces state of the TI is probed at high *f*. At low frequency (Fig. 4(a)) the induced current circulates in the bulk of the sample. In Fig. 4(c) we plot the pickup voltage versus sample surface area for all our five samples having varied thickness. Note that above 120 K at 5 kHz as the current induces in the bulk of the sample, the pickup signal saturates to a value close to 1 mV for all samples, which shows no scaling with surface area. In fact, at high frequency and low *T* (65 kHz and 25 K) as the highly conducting surface states are effectively probed, the pickup voltage scales with the sample surface area. At low *T*, numerous studies show [13,14,21,22] as well as SdH oscillation in Fig. 1(a) confirm the dominance of surface conductivity in the TI. In this regime the observation of $V(f) \propto f^{\alpha}$, where $\alpha \sim 1$, suggests the linear frequency dependence of pickup voltage *V(f)*, is related to contribution to electrical conduction from the surface state of the TI. The emergence of the $V(f) \propto f^2$ at low frequency and higher *T* is related to the bulk contribution to conductivity where *V(T)* shows thermally activated behavior.

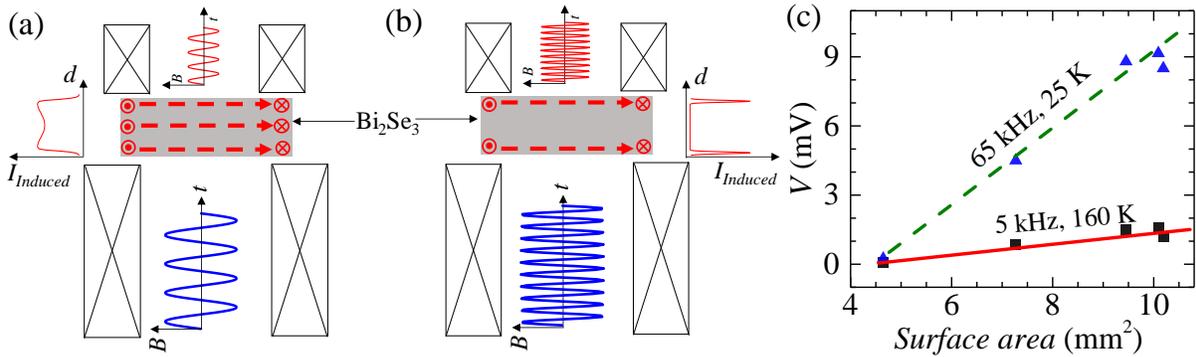

FIG. 4. (a) Shows the schematic of the magnitude of the induced the current distribution across the sample cross-section at a lower frequency. The induced current is shown to flow across the entire volume of the sample. The AC excitation magnetic field in the primary is shown in blue while the field induced in the pickup coils through the currents induced in the sample is shown in red. A schematic plot of the induced current profile across the sample cross-section

at different *f* is shown. (b) Shows at a higher frequency, the nature of the induced current distribution which flows only in TI's high conducting surface states. The nature of the excitation field in the excitation coil (blue) and the induced field in the pickup coil (red) are also shown. (c) Shows the pickup voltage response with sample surface area at 5 kHz, 160 K and 65 kHz, 25 K.

At low $T$ our transport measurements in Fig. 1 as well as past observations [13,14,21,22] have confirmed the dominance of surface contribution to conductivity at low $T$ in the TI. In this low $T$ regime our pickup signal exhibits a linear frequency dependent regime. This suggests that the linear frequency dependence of the pickup signal is related to the surface conducting state in the TI. Similarly, at high $T$ the observation from our transport measurements and those of other [21,22] of the emergence of bulk conducting response, coincides with the observed quadratic dependence at low frequencies of the pickup signal. Therefore, this quadratic frequency dependence at low frequency of the pickup signifies the emergence of bulk contribution to electrical conductivity. Another possible explanation to the observed frequency dependence is given as follows: The pickup voltage develops due to the current induced in the sample. These induced currents depend on the electrical conductivity of the sample. Based on the above discussion the electrical conductivity of the TI has two contributions $\sigma_b(T) = \sigma_{b0} \exp\left(-\Delta/K_b T\right)$ and $\sigma_s(T) = 1/(C+DT)$. The pickup voltage is given by (see S9 section in supplementary information of Ref. [46] for detailed calculation)

$$V_{pickup}(\omega) = \xi(r,z)\Phi\omega^2 e^{-i\omega t}\left[\frac{\sigma_{os}}{C+DT} + \frac{\sigma_{ob}}{e^{\Delta/KT}}\right]\frac{1}{1+i\omega\tau}, \quad (1)$$

where $\xi(r, z)$ is a geometric factor which is a function of the pickup coil radius ($r$), the height between the two coils ($z$), $\Phi$ is the magnetic flux passing through the sample, $\tau$ is scattering time scale and $\omega = 2\pi f$. Terahertz spectroscopy measurements [49,50] show that bulk and surface have different value of scattering time scale $\tau$. Hence, we assume two different limits in eqn. (1) (see S9 of Ref. [46] for details). When bulk state dominates in conduction, we use a limit $f\tau \ll 1$. Under this assumption Eq. (1) gives us

$$V_{pickup}(f) \approx \Phi_0 \xi(r,z) f^2 e^{-i2\pi ft}\sigma_b. \quad (2)$$

If the surface state start to dominate in conduction, using the limit $f\tau \gg 1$, Eq. (1) shows $V(f) \propto f^{\alpha}$, viz.,

$$V_{pickup}(f) \approx \Phi_0 \xi(r,z) e^{-i2\pi ft} \sigma_s f. \tag{3}$$

Figure 5(a) shows the $V(T)$ measurements of Fig. 3(b) for the S69 sample. The data is fitted with the expression:

$$V(T) = P_{surface}(\text{Eq. 3}) + P_{bulk}(\text{Eq. 2}), \tag{4}$$

where, $P_{surface}$ and $P_{bulk}$ are mean fractions of the surface and bulk contribution to the pickup voltage with $P_{surface} + P_{bulk} = 1$. Note the $T$ dependence of $\sigma_s$ and $\sigma_b$ has already been shown by the black line and red dashed line in Fig. 3(b), respectively. The S69 data in Fig. 5(a) fits with $P_{surface} = 35\%$, $P_{bulk} = 65\%$. Note that at 65 kHz, the $P_{surface}$ is sensitive to the thin conducting regions within the sample where shielding currents are induced by the AC field. The minima in $V(T)$ at 70 K (Fig. 5(a)) suggests regions in the TI which exhibited predominantly surface conductivity response at low $T$ (Fig. 3(c)), with increasing $T$ begins to display bulk conductivity. The surface contribution is gradually degraded by transport channel being created in the bulk, as a result we see only an average 35% contribution to the total conductivity from the surface.

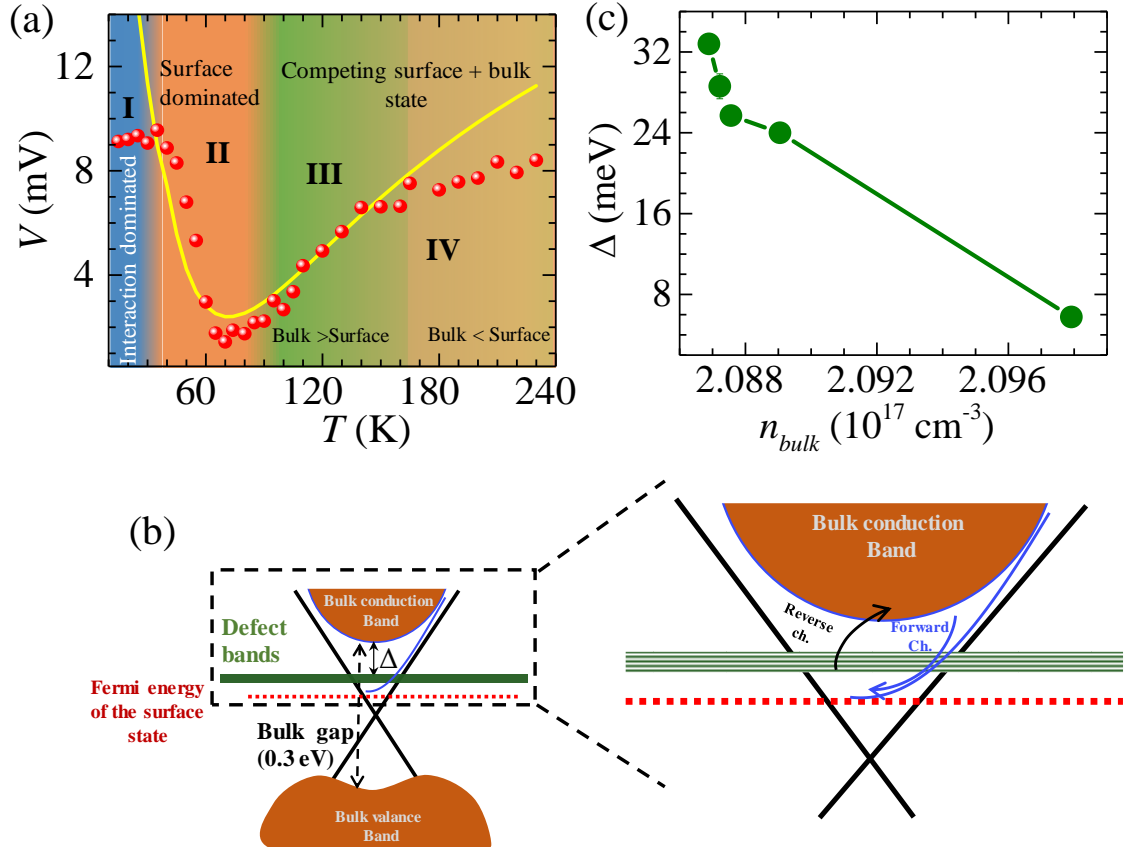

FIG. 5. (a) The *V(T)* response for sample S69 is shown in red circles for 65 kHz frequency. The solid yellow line is the fit to the data, done using Eq. **4**. Four distinct regions (I-IV) are shown with four different colors (see text for details). (b) A schematic showing the bulk conduction and valence bands as well as the surface states. The schematic on the right is an expanded portion of the left schematic. The violet curve in the right schematic shows upward band bending of the conducting band towards the Fermi energy of the surface state due to Se vacancies. The forward channel (Forward Ch.) is associated with electrons migrating from the bulk to the surface, the reverse channel (Reverse Ch.) is associated with electrons being thermally activated from the surface into the bulk. Black lines show the Dirac cone associated with the surface states. The Fermi energy is shown by red dotted line. The green region represents the defect states produced by the Se vacancies. The activation energy gap ($\Delta$) is between the defect state and minima of the conduction band (c) Activation energy gap ($\Delta$) calculated from our model is plotted as a function of bulk carrier ($n_{bulk}$). The $n_{bulk}$ is calculated using, $n_{bulk} = \sigma_b / \mu(d)e$, where $\mu(d) = \frac{3000}{1+140/d}$ [23] is the thickness (*d*) dependent mobility and $\sigma_b \sim 10^3$ S/m (based on the value used in the simulations above).

The fitting in Fig. 5(a) shows that at high $T$ above 170 K there is a deviation from the Eq. **4** fit and the $V(T)$ becomes weakly temperature dependent. We have found in all the samples, the $V(f)$ at high temperatures exhibits linear and quadratic frequency dependence at high and low frequencies respectively (for eg. see 150 K data in Fig .3(a)). This behavior suggests the presence of surface conductivity coexisting with a bulk conducting regime in the sample even at high temperature. The temperature dependence of $V(T)$ below 70 K in Figs. 3(b) and 5(a) suggests that the electrical conductivity of the sample obeys $\sigma_s(T) = 1/(C+DT)$. This form of temperature dependence of the electrical conductivity has been reported for surface conducting state in TI [21,22]. The $T$ dependent part of the electrical conductivity, $\sigma(T) = 1/(C+DT)$, sets in above 40 K (i.e., $D \gg C$). Our, and other investigations [51,52] shows the presence of a temperature independent conductivity regime below 40 K, which is a feature of metallic conductivity of the topologically protected surface states in TI materials. Earlier studies in $Bi_2Se_3$ studies show that the $T$ dependent conductivity above 40 K is due to electron phonon interactions [51]. With lowering of $T$ (below 40 K) where the surface conductivity of the TI dominates, the electron – phonon coupling significantly weakens [12]. From the above expression the $T$ independent conductivity regime is obtained when $D \ll C$ at $T < 40$. In this temperature regime where conductivity is via the topological surface states, magneto-transport studies show weak antilocalization due to the chiral nature of the currents [21,33,53]. In this regime studies suggest [33] electron interactions play a role in determining the behavior of conductivity of the TI materials. Hence below 40 K where there is a saturated behaviour of conductivity, the effect of electron – phonon interaction may be relatively diminished compared to that of electron interaction effects. In view of this we identify an electron interaction dominated regime at low $T$.

Based on our above discussions, in Fig. 5(a) we identified four different regimes: (i) Region I (below ~ 40 K the strong interaction dominated surface regime): In this regime, surface conductivity dominates in the TI and conductivity is nearly temperature independent due to strong electron-electron interaction effects (ii) Region II (temperature dependent surface state): Here the surface state dominates the conductivity of the TI with a temperature dependence of the form $\sigma_s(T) = 1/(C+DT)$ due to the onset of electron-phonon scattering. As the temperature increases, the surface conductivity falls until it reaches close to the bulk conductivity value, which is close to the minima in *V(T)*. (iii) Region III (bulk dominated state): In this region, which extends about 70 K to 170 K, Fig. 3(b) shows a good fit to bulk behavior (see dashed red curve in Fig. 3(b)). However, the fit doesn't imply that in region III the surface contribution to conductivity is zero. Figure 3(b) shows that in regime III, the surface contribution to conductivity (solid black line) is much smaller than the bulk contribution, due to which the fit to bulk appears to be good and bulk dominated. Figure. 3(a) shows that for temperatures in regime III (like 90 K and 150 K) the *V(f)* has quadratic as well a nearly linear regime ($V(f) \propto f^\alpha$, where $\alpha \sim 1$), which shows the presence of both bulk and surface contributions to conductivity. Hence, region III is a regime with the bulk contribution to conductivity dominating over a weak surface contribution. Here the response from the bulk in the TI shows a thermally activated conductivity behavior. The typical thermal activation energy scale is estimated to be ~ 6 meV - 30 meV by studying five different samples. (iv) Region IV: This is an unusual regime found at high temperatures, where *V(T)* deviates from Eq. (4) fit. In this regime, we recall similar linear *V(f)* behavior (inset Fig. 3(a)). Just like the weakly *T* dependent surface conductivity regime is present at low *T*, we propose the weakly temperature independent regime at high *T* is where the effects of surface conductivity of a TI become important once again. We would like to reiterate here that surface and bulk, both conductivities contribute to the pickup voltage, as the induced currents are generated in both surface and bulk of the sample. The region III and region IV (Fig. 5(a)) are associated with a competition between bulk and surface state contributions to conductivity. The change in curvature of *V(T)* curve in Fig. 5(a) is going from region III to region IV which signifies a competition between surface and bulk contributions to conductivity. In (region III) the thermally activate nature of transport we have showed that bulk conductivity dominates. Recall inset of Fig. 3(a) at high *T*, showed a nearly linear frequency dependence, viz., $V(f) \propto f^\alpha$, where $\alpha \sim 1$ when *T* is in region IV. This suggest the re-emergence of significant surface contributions to conductivity along with bulk. Here within the temperature window III the bulk contribution to conductivity dominates over the surface contribution, while in region IV it is vice versa.

## VI.   Simulations of the mutual inductance measurements:

We simulate the electromagnetic response of this topological insulator with defects using COMSOL multi-physics software. In Fig. 6(a) inset, we first model a defect free TI with high conducting thin metallic sheets (yellow) sandwiching a bulk low conducting slab (blue). We solve the Maxwell equation for the modelled TI subjected to a time varying magnetic field applied using coil parameters identical to our experiment (see S2 section in supplementary information of Ref. [46]). We use an AC excitation current of $I = 150$ mA (similar to our experiment). First, the rate of change of induced flux ($\dot{\varphi} = M\dot{I}$) in the pickup coil is calculated from which the mutual inductance (real and imaginary components) $M'$ and $M''$ are calculated (analog to our experiments, $M'' \ll M'$). The modelled TI has crystal thickness and surface area similar to S20 sample, with a thickness of 5 nm for the metallic high conducting sheets and 20 µm for the bulk insulating layer. The above is based on estimates indicating the thickness of conducting surface in $Bi_2Se_3$ is between 5 to 10 nm [54,55]. In Fig. 6(a) we see the data begins to look like the experimental $V(f)$ data when the bulk conductivity $\sigma_b \sim 10^3$ S/m and the conductivity of the metallic sheets $\sigma_s \sim 10^{11}$ S/m (note OFHC triple nine purity copper has conductivity $\sim 10^8$ S/m). Earlier transport measurements have shown conductivity of surface states in $Bi_2Se_3 \sim 10^4$ S/m to $10^5$ S/m [22,56]. It may be mentioned here that in conventional transport measurements due to inability to distinguish between bulk and surface conducting channels, the measured surface conductivity via transport measurements could be much lower than the actual value. (Details on the behavior of the skin depth and its estimate is shown in the supplementary section S10 of Ref. [46]). Note that the relation between $M'$ and induced voltage in the pickup coil is $V' \propto \omega M'$ (as $M'' \ll M'$ or $V'' \ll V'$, $M'(f) = \frac{V'(f)}{\omega I}$ where $I = 150$ mA, $\omega = 2\pi f$). Figure 6(a) shows simulated $M'$ as a function of frequency for different conductivities of the TI surface while keeping the bulk conductivity constant at $\sim 10^3$ S/m. The value $\sigma_b$ we use is based on estimates in literature [35]. The general behavior of $M'(f)$ is that, it increases linearly with frequency and it saturates at higher frequency. At lower frequency regime where $M'(f) \propto f$ (see dotted line drawn in Fig. 6(a)) corresponds the pickup voltage, $V(f) \propto f^2$. At higher $f$ where the $M'$ is weakly dependent on $f$, corresponds to $V(f) \propto f$. Figure 6(a) shows that the $M'$ begins to show saturating features as the surface state conductivity ($\sigma_s$) increases. This confirms that highly conducting surface states are responsible for the nearly linear frequency dependence of pickup voltage $V(f)$ in the high frequency regime (see Fig. 3). Note that while the simulations in Fig. 6(a) shows the $M'(f)$ saturates above 40 kHz, however in our experiments (Fig. 3) showed that $V(f) \propto f^{\alpha}$ (where $\alpha \sim 1$) regime (i.e., $M'(f) \sim$ constant) sets in above 20 kHz and not from 40 kHz.

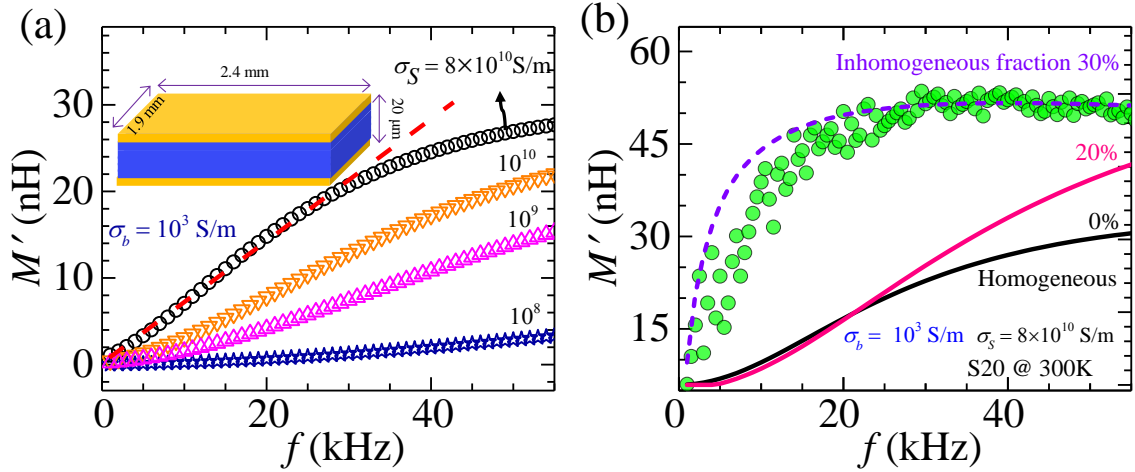

FIG. 6. Inset (a) shows the schematic of an ideal TI. Blue color represents bulk state and yellow color represents high conducting surface state. The thickness of the high conducting surface state is taken as 5 nm (thickness of the yellow region), the other relevant dimensions are height = 20 μm, width = 1.9 mm, and length= 2.4 mm. The main panel shows the behavior of the simulated $M'(f)$ for different conductivity of surface state keeping bulk conductivity constant (~$10^3$ S/m). Red dash line shows the linear dependence of $M'$. It is clear that $M'$ seems to saturate beyond 40 kHz whereas the experimentally measured data shows saturation from 20 kHz onwards. (b) Shows the $M'$ is simulated as a function of $f$ for different inhomogeneity levels (0%, 20% and 30%) by incorporating conducting channels inside the bulk (see text for details and section S2 of supplementary information, of Ref. [46]). Comparison of simulation and experimental result (scattered data) of $M'$ is shown for S20 sample at room temperature.

In order to match our simulations with the observed saturation of $M'(f)$ at relatively lower $f$ (from 20 kHz), without significantly enhancing the surface conductivity $\sigma_S$ (far beyond what have been reported in literature for these samples), we consider a situation where conducting channels exist in the bulk which connect to the high conducting surface states. We argue later these high conducting channels in the bulk are generated by Se vacancies. Such a state of the TI we refer to as an inhomogeneous topological insulator state. We incorporate inhomogeneity in our model for the TI by considering a distribution of solid high conductivity cylindrical regions threading through the bulk of the sample connecting the top and bottom high conducting surface sheets of the TI (see section S2 in the supplementary information of Ref. [46] for a schematic of the modelled inhomogeneous TI system). In our simulation the percentage inhomogeneity in the TI is varied by changing the density of cylinders and the conductivity of the cylinders is taken as $\sigma_s$. Figure 6(b) shows that the results of simulations match with our experimental results by, using $\sigma_s \sim 10^{11}$ S/m, $\sigma_b \sim 10^3$ S/m and 30% inhomogeneity. The introduction of the 30% inhomogeneity in our TI model causes the $M'(f)$ to saturate from lower frequencies of 20 kHz. The presence of 30% inhomogeneity in the sample would affect the net bulk electrical conductivity. However, estimating the extent of change in the net bulk electrical conductivity isn't clear as yet as this would statistically depend on the fraction of high conductivity and low channels contributing to the electrical conduction paths when electric current flows through the bulk. We find that our above simulation results are not sensitive to the details of the spatial distribution of inhomogeneity introduced in the TI, but rather the result are quite sensitive on the inhomogeneity fraction in the TI (see section S2 in the supplementary information of Ref. [46]). Figure 6(b) shows the close match between the simulated $M'(f)$ (purple dashed line) and the measured voltage data (green scattered data) for S20 sample at 300 K using the inhomogeneous TI model. Since in this high frequency regime the skin depth in the TI is estimated to be ~100 nm (see the supplementary information, S10 of Ref. [46]), we suggest that the inhomogeneity discussed above are clustered over length scales which are ≤ 100 nm (in our model we take the spatial extend of inhomogeneity in the TI as the diameter of the cylinders). Due to the presence of parallel channels of electrical conduction between the surface and bulk of the TI material, it is difficult to identify the surface conductance and inhomogeneity through the bulk in a transport measurement, especially at intermediate temperature regimes where both contributions are admixed. However, the analysis of SdH oscillations at different $T$ help to give a signature to the onset of the bulk contribution to electrical conductivity.

Analysis of the SdH oscillations in our Bi$_2$Se$_3$ (see Fig.1 and its caption for details) using the Lifshitz-Kosevich (LK) equation [35,57,58] gives a frequency of the SdH oscillations ($F$) = 46.95 ± 0.25 T. As $F = (4\pi^2 \hbar n_s)/e$, where $n_s$ is the surface carrier density, we get $n_s$ = (2.268 ± 0.012) cm$^{-2}$, which compares well with earlier estimates in this batch of samples [35]. Using $n_s$ we estimate the Fermi wave vector for the 2D surface states in our sample is, $k_f = \sqrt{2\pi n_s}$ = (0.0377 ±0.0027) Å$^{-1}$. By comparing this $k_F$ value with the ARPES spectrum measured for Bi$_2$Se$_3$ samples (see Fig.1 of Ref. [59]), we see that the Fermi energy ($E_f$) should be located approximately 30 meV above the Dirac point and about 100 meV below the bottom of the bulk conduction band (this comparison is illustrated in section S12 of supplementary information, Ref. [46]). If we assume the origin of the SdH oscillation in the transport measurements is due to the bulk electrons, then bulk carrier density should be $k_F^2/4\pi c = 3.8 \times 10^{18}$ cm$^{-3}$, where c=28.64 Å is the Bi$_2$Se$_3$ lattice spacing along c axis. However, the bulk carrier density ($n_{bulk}$) in the TI material as estimated from the Hall coefficient (Fig. 1(b)), is $2.83 \times 10^{19}$ cm$^{-3}$. This comparison shows that the features of our transport data at low $T$ cannot be reconciled with bulk conduction electrons contributing to conductivity rather they are arising from 2D surface electrons. The SdH oscillations in the transport measurements (see Fig. 1) analyzed using the Lifshitz-Kosevich (LK) equation also reveal information on the Berry phase, $\phi = 2\pi\beta$ (see Fig. 1(b) inset). For the two-dimensional Dirac fermion in TI $\phi = \pi$ suppresses the back scattering from disorder due to destructive interference. From the analysis of the SdH at different $T$, we obtain a value close to $\pi$ at low $T$ below 10 K (see Fig.1(b) inset). Thus at low $T$ the observed value of $\phi$ close to $\pi$ suggests that the SdH oscillation arises from 2D surface Dirac electrons. As bulk contributions to conductivity increases with $T$, the SdH oscillations weaken and disappear (see Fig. 1(a)) and the $\phi$ value significantly deviates from $\pi$ above 10 K (see Fig.1(b) inset). Therefore, at low $T$ the transport data shows conductivity of the TI is surface dominated which gets progressively affected as the bulk conductivity sets in at higher $T$. However the temperature range over which the above analysis can be done is very limited. While the above observations are consistent with our results based on pickup coil measurements, however our technique is applicable over a wider $T$ range.

### VII. Discussion on the role of disorder in Bi$_2$Se$_3$:

A bulk gap in Bi$_2$Se$_3$ of 0.3 eV suggests that thermally activated conductivity should be seen at significantly elevated temperatures, but we see activated conductivity from $T \sim 70$ K onwards (viz., the region III in Fig. 5(a)). In Bi$_2$Se$_3$ samples it is known Se vacancies add defect states in the bulk. These defects states add excess electrons in the bulk due to which the conductivity of the bulk increases. The measured Δ values are almost an order of magnitude smaller than the estimates of bulk band gap in Bi$_2$Se$_3$ of about 0.3 eV [7,10, 26,27]. One possible scenario could be, the gap Δ corresponds to a gap between the defect states generated by Se vacancies and bottom of conduction band (see Fig. 5(b)). It is known that Se vacancy causes an upward band bending (Fig. 5(b)) of the states near the bottom of the conduction band towards the surface state [60]. The band bending favors the migration of electrons generated in the bulk (due to Se vacancies) to the surface. We refer to this as the forward channel (see Fig. 5(b)). Above 70 K the thermal energy is sufficient for thermal activation of charges from defects states into the bulk. We refer to this as the reverse channel. This reverse channel for charge migration begins at $T > \Delta/k_B$ (Fig. 5(b)). Recall, that in $Bi_2Se_3$ excess charge carriers are produced by Se vacancies. In Fig. 5(c) we estimated the behavior of carrier density in the bulk ($n_{bulk}$) using a known form [23], $n_{bulk} = \frac{\sigma_b}{\mu(d)e}$, where $\mu(d) = \frac{3000}{1 + 140/d}$ is the thickness ($d$) dependent mobility and $\sigma_b \sim 10^3$ S/m. Thus the activation barrier $\Delta$ is larger for smaller bulk carrier density. Akin to a tight binding like picture, we propose the hopping back and forth of charges between the defect states in the bulk and surface states enhances the average kinetic energy of charges leading to the formation of a broad defect band and this also helps to open a gap $\Delta$ near the surface states. The $\Delta$ decreases with $n_{bulk}$ as seen in Fig. 5(c), could be a result of enhanced Coulombic interactions in these lower electron doped samples suppressing the kinetic energy and localizing the charges. Enhanced interaction effects lead to the weak $T$ dependence of conductivity not only at low $T$ below 40 K but also at high $T$ above 180 K (Fig. 5(a)). Thus Selenium vacancy induced electron doping of the bulk leads to an interplay between bulk and surface conductivity, leading to an inhomogeneous TI state in $Bi_2Se_3$. Based on the above discussion, we propose that in region III of Fig. 5(a) the bulk contribution to the conductivity dominates over the surface, due to thermal excitation of charge carrier in the bulk state. However, at higher temperature (region IV), the excess charge migration to surface state due to Se vacancies results in the contribution of the surface conductivity to enhance once again. It is due to this effect, we see the experimental data deviating from the fitting line in high temperature regime (region IV in Fig. 5(a)). We believe more detailed work using techniques like STM and ARPES, is required to probe the effect of defect states lying within the bulk and the resulting gap emerging in the TI.

### VIII. Conclusions:

In conclusion, our non-contact two coil mutual inductance measurements in $Bi_2Se_3$ single crystals suggest that the surface states are coupled to bulk state with filamentary high conducting structures. The inhomogeneous state gets precipitated in the bulk of the TI only above the threshold temperature, which is associated with the activation of charge between the surface and bulk. This leads to an interplay between bulk and surface dominated transport regimes at different $T$, which is produced as a result of disorder introduced in the system. While material disorder leads to interplay between bulk and surface conductivity in a TI material, it is also responsible for resurfacing of the surface conductivity at high temperature which is a potentially feature useful for applications.

**Acknowledgements**

SSB acknowledges the funding support from DST (TSDP), IIT Kanpur for funding support.

**References**

1. Joel E. Moore, The birth of topological insulators, Nature **464**, 194 (2010).
2. M. Z. Hasan and C. L. Kane, Colloquium: Topological insulators, Rev. Mod. Phys. **82**, 3045 (2010).
3. D. Hsieh, D. Qian, L. Wray, Y. Xia, Y. S. Hor, R. J. Cava and M. Z. Hasan, A topological Dirac insulator in a quantum spin Hall phase, Nature **452**, 970 (2008).
4. L. Fu, C. L. Kane and E. J. Mele, Topological Insulators in Three Dimensions, Phys. Rev. Lett. **98**,106803 (2007).
5. D. Hsieh, Y. Xia, L. Wray, D. Qian, A. Pal, J. H. Dil, J. Osterwalder, F. Meier, G. Bihlmayer, C. L. Kane, Y. S. Hor, R. J. Cava and M. Z. Hasan, Observation of Unconventional Quantum Spin Textures in Topological Insulators, Science **323**, 919-922 (2009).
6. Lukas Zhao, Haiming Deng, Inna Korzhovska, Zhiyi Chen, Marcin Konczykowski, Andrzej Hruban, Vadim Oganesyan and Lia Krusin-Elbaum, Singular robust room-temperature spin response from topological Dirac fermions, Nat. Mate. **13**, 580–585 (2014).
7. Y. Xia1, D. Qian, D. Hsieh, L. Wray, A. Pal, H. Lin, A. Bansil, D. Grauer, Y. S. Hor, R. J. Cava and M. Z. Hasan, Observation of a large-gap topological-insulator class with a single Dirac cone on the surface, Nat. Phys. **5**, 398-402 (2009).
8. Y. L. Chen, J. G. Analytis, J.-H. Chu, Z. K. Liu, S.-K. Mo, X. L. Qi, H. J. Zhang, D. H. Lu1, X. Dai, Z. Fang, S. C. Zhang, I. R. Fisher, Z. Hussain and Z.-X. Shen, Experimental Realization of a Three-Dimensional Topological Insulator $Bi_2Te_3$, Science **325**, 178-181 (2009).
9. Peng Cheng, Canli Song, Tong Zhang, Yanyi Zhang, Yilin Wang, Jin-Feng Jia, Jing Wang, Yayu Wang, Bang-Fen Zhu, Xi Chen, Xucun Ma, Ke He, Lili Wang, Xi Dai, Zhong Fang, Xincheng Xie, Xiao-Liang Qi, Chao-Xing Liu, Shou-Cheng Zhang and Qi-Kun Xue, Landau Quantization of Topological Surface States in $Bi_2Se_3$, Phys. Rev. Lett. **105**, 076801 (2010).
10. H. Zhang, C. X. Liu, X. L. Qi, X. Dai, Z. Fang and S. C. Zhang, Topological insulators in $Bi_2Se_3$, $Bi_2Te_3$ and $Sb_2Te_3$ with a single Dirac cone on the surface, Nat. Phys. **5**, 438-442 (2009).

11. Zhuojin Xie, Shaolong He, Chaoyu Chen, Ya Feng, Hemian Yi, Aiji Liang, Lin Zhao, Daixiang Mou, Junfeng He, Yingying Peng, Xu Liu, Yan Liu, Guodong Liu, Xiaoli Dong, Li Yu, Jun Zhang, Shenjin Zhang, Zhimin Wang, Fengfeng Zhang, Feng Yang, Qinjun Peng, Xiaoyang Wang, Chuangtian Chen, Zuyan Xu and X. J. Zhou, Orbital-selective spin texture and its manipulation in a topological insulator, Nat. Comm. **5**, 3382-9 (2014).

12. Z.-H. Pan, A. V. Fedorov, D. Gardner, Y. S. Lee, S. Chu, and T. Valla, Measurement of an Exceptionally Weak Electron-Phonon Coupling on the Surface of the Topological Insulator $Bi_2Se_3$ Using Angle-Resolved Photoemission Spectroscopy, Phys. Rev. Lett. **10**, 187001 (2012).

13. D. X. Qu, Y. S. Hor, J. Xiong, R. J. Cava and N. P. Ong, Quantum Oscillations and Hall Anomaly of Surface States in the Topological Insulator $Bi_2Te_3$, Science **329**, 821-824 (2010).

14. L. A. Jauregui, M. T. Pettes, L. P. Rokhinson, L. Shi and Y. P Chen, Gate Tunable Relativistic Mass and Berry's phase in Topological Insulator Nanoribbon Field Effect Devices, Sci. Rep. **5**, 8452 (2015).

15. Pedram Roushan, Jungpil Seo, Colin V. Parker, Y. S. Hor, D. Hsieh, Dong Qian, Anthony Richardella, M. Z. Hasan, R. J. Cava and Ali Yazdani, Topological surface states protected from backscattering by chiral spin texture, Nature **460**, 1106 (2009).

16. X. L. Qi and S. C. Zhang, Topological insulators and superconductors, Rev. Mod. Phys. **83**, 1057-1110 (2011).

17. H. Soifer, A. Gauthier, A. F. Kemper, C. R. Rotundu, S.-L. Yang, H. Xiong, D. Lu, M. Hashimoto, P. S. Kirchmann, J. A. Sobota and Z.-X. Shen, Band-Resolved Imaging of Photocurrent in a Topological Insulator, Phys. Rev. Lett. **122**,167401 (2019).

18. S. A. Wolf, D. D. Awschalom, R. A. Buhrman, J. M. Daughton, S. von Molnár, M. L. Roukes, A. Y. Chtchelkanova and D. M. Treger, Spintronics: a spin-based electronics vision for the future, Science **294**, 1488 (2001).

19. D. K. Kim, Y. Lai, B. Diroll, C. Murray and C. F. Kagan, Flexible and low-voltage integrated circuits constructed from high-performance nanocrystal transistors, Nat. Comm. **3**, 1216-6 (2012).

20. P. J. Leek, J. M. Fink, A. Blais, R. Bianchetti, M. Göppl, J. M. Gambetta, D. I. Schuster, L. Frunzio, R. J. Schoelkopf and A. Wallraff, Observation of Berry's phase in a solid-state qubit, Science **318**, 1889 (2007).

21. B. F. Gao, P. Gehring, M. Burghard and K. Kern, Gate-controlled linear magnetoresistance in thin $Bi_2Se_3$ sheets, Appl. Phys. Lett. **100**, 212402 (2012).

22. Yang Xu, Ireneusz Miotkowski, Chang Liu, Jifa Tian, Hyoungdo Nam, Nasser Alidoust, Jiuning Hu, Chih-Kang Shih, M. Zahid Hasan and Yong P. Chen, Observation of topological surface state quantum Hall effect in an intrinsic three-dimensional topological insulator, Nat. Phys. **10**, 956 (2014).

23. Yong Seung Kim, Matthew Brahlek, Namrata Bansal, Eliav Edrey, Gary A. Kapilevich, Keiko Iida, Makoto Tanimura, Yoichi Horibe, Sang-Wook Cheong and Seongshik Oh, Thickness-dependent bulk properties and weak antilocalization effect in topological insulator $Bi_2Se_3$, Phys. Rev. B **84**, 073109 (2011).

24. Diptasikha Das, K. Malik, S. Bandyopadhyay, D. Das, S. Chatterjee and A. Banerjee, Magneto-resistive property study of direct and indirect band gap thermoelectric Bi-Sb alloys, Appl. Phys. Lett. **105**, 082105 (2014).

25. J. G. Analytis, R. D. McDonald, S. C. Riggs, J. H. Chu and G. S. Boebinger, Two-dimensional surface state in the quantum limit of a topological insulator, Nat. Phys. **6**, 960–964 (2010).

26. P. Larson, V. A. Greanya, W. C. Tonjes, R. Liu, S. D. Mahanti and C. G. Olson, Electronic structure of $Bi_2X_3$ (X=S, Se, T): compounds: Comparison of theoretical calculations with photoemission studies, Phys. Rev. B **65**, 085108-11 (2001).
27. I. A. Nechaev, R. C. Hatch, M. Bianchi, D. Guan, C. Friedrich, I. Aguilera, J. L. Mi, B. B. Iversen, S. Blügel, Ph. Hofmann, and E. V. Chulkov, Evidence for a direct band gap in the topological insulator $Bi_2Se_3$ from theory and experiment, Phys. Rev. B **87**, 121111-5 (2013).
28. G. R. Hyde, H. A. Beale, I. L. Spain and J. A. Woollam, Electronic properties of $Bi_2Se_3$ crystals, J. Phys. Chem. Solids. **35**, 1719 (1974).
29. H. M. Benia, C. Lin, K. Kern and C. R. Ast, Reactive Chemical Doping of the $Bi_2Se_3$ Topological Insulator, Phys. Rev. Lett. **107**, 177602-4 (2011).
30. J. M. Zhang, W. Ming, Z. Huang, G. B. Liu, X. Kou, Y. Fan, K. L. Wang and Y. Yao, Electronic and magnetic properties of the magnetically doped topological insulators $Bi_2Se_3$ $Bi_2Te_3$, and $Sb_2Te_3$, Phys. Rev. B **88**, 235131-9 (2013).
31. Nicholas P. Butch, Kevin Kirshenbaum, Paul Syers, Andrei B. Sushkov, Gregory S. Jenkins, H. Dennis Drew and Johnpierre Paglione, Strong surface scattering in ultrahigh-mobility $Bi_2Se_3$ topological insulator crystals, Phys. Rev. B **81**, 241301-4 (2010).
32. H. D. Li, Z. Y. Wang, X. Kan, X. Guo, H. T. He, Z. Wang, J. N. Wang, T. L. Wong, N. Wang and M. H. Xie, The van der Waals epitaxy of $Bi_2Se_3$ on the vicinal Si (111) surface: an approach for preparing high-quality thin films of a topological insulator, New. J. Phys. **12**, 103038 (2010).
33. Minhao Liu, Cui-Zu Chang, Zuocheng Zhang, Yi Zhang, Wei Ruan, Ke He, Li-li Wang, Xi Chen, Jin-Feng Jia, Shou-Cheng Zhang, Qi-Kun Xue, Xucun Ma and Yayu Wang, Electron interaction-driven insulating ground state in $Bi_2Se_3$ topological insulators in the two-dimensional limit, Phys. Rev. B **83**, 165440-6 (2011).
34. K. Hoefer, C. Becker, C. D. Rata, J. Swanson, P. Thalmeier and L. H. Tjeng, Intrinsic conduction through topological surface states of insulating $Bi_2Te_3$ epitaxial thin films, Proc. Natl. Acad. Sci. **A111**, 14979−14984 (2014).
35. T. R. Devidas, E. P. Amaladass, Shilpam Sharma, R. Rajaraman, D. Sornadurai, N. Subramanian, Awadhesh Mani, C. S. Sundar and A. Bharathi, Role of Se vacancies on Shubnikov-de Haas oscillations in $Bi_2Se_3$: A combined magneto-resistance and positron annihilation study, Euro Phys. Lett. **108**, 67008 (2014).
36. E. P. Amaladass, T. R. Devidas, S. Sharma, C. S. Sundar, A. Mani and A. Bharathi, Magneto-transport behaviour of $Bi_2Se_{3-x}Te_x$: role of disorder. J. Phys: Condens. Matter **28**, 075003 (2016).
37. C. E. ViolBarbosa, Chandra Shekhar, Binghai Yan, S. Ouardi, Eiji Ikenaga, G. H. Fecher and C. Felser, Direct observation of band bending in the topological insulator $Bi_2Se_3$, Phys. Rev. B **88**,195128 (2013).
38. Desheng Kong, Judy J. Cha, Keji Lai, Hailin Peng, James G. Analytis, Stefan Meister, Yulin Chen, Hai-Jun Zhang, Ian R. Fisher, Zhi-Xun Shen and Yi Cui, Rapid Surface Oxidation as a Source of Surface Degradation Factor for $Bi_2Se_3$, ACS Nano **5**, 4698 (2011).
39. Mintu Mondal, Anand Kamlapure, Madhavi Chand, Garima Saraswat, Sanjeev Kumar, John Jesudasan, L. Benfatto, Vikram Tripathi, and Pratap Raychaudhuri, Phase Fluctuations in a Strongly Disordered s-Wave NbN Superconductor Close to the Metal-Insulator Transition, Phys. Rev. Lett. **106**, 047001-4 (2011).
40. Anand Kamlapure, Mintu Mondal, Madhavi Chand, Archana Mishra1, John Jesudasan, Vivas Bagwe, L. Benfatto, Vikram Tripathi and Pratap Raychaudhuri, Measurement of magnetic penetration depth and superconducting energy gap in very thin epitaxial NbN films, Appl. Phys. Lett. **96**, 072509 (2010).

41. Ming-Chao Duan, Zhi-Long Liu, Jian-Feng Ge, Zhi-Jun Tang, Guan-Yong Wang, Zi-Xin Wang, Dandan Guan, Yao-Yi Li, Dong Qian, Canhua Liu and Jin-Feng Jia, Development of in situ two-coil mutual inductance technique in a multifunctional scanning tunneling microscope, Rev. Sci. Instrum. **88**, 073902 (2017).

42. Indranil Roy, Prashant Chauhan, Harkirat Singh, Sanjeev Kumar, John Jesudasan, Pradnya Parab, Rajdeep Sensarma, Sangita Bose and Pratap Raychaudhuri, Dynamic transition from Mott-like to metal-like state of the vortex lattice in a superconducting film with a periodic array of holes, Phys. Rev. B **95**, 054513 (2017).

43. J. H. Claassen, M. L. Wilson, J. M. Byers, and S. Adrian, Optimizing the two-coil mutual inductance measurement of the superconducting penetration depth in thin films, J. Appl. Phys. **82**, 6, (1997).

44. Stefan J. Turneaure, Eric R. Ulm, and Thomas R. Lemberger, Numerical modeling of a two-coil apparatus for measuring the magnetic penetration depth in superconducting films and arrays, J. Appl. Phys. **79**, 4221 (1996).

45. H. Hochmuth and M. Lorenz, Inductive determination of the critical current density of superconducting thin films without lateral structuring, Physica C **220**, 209-214 (1994).

46. See Supplemental Material at " Insert the URL" for S1 section deals with the coil details; S2 section shows the differential equations and geometry of the measurement configuration which are used to solve the differential equations by the Comsol software; S3 section shows the background and pure Cu sheet pickup signal; S4, S5, S6, S7 and S8 sections show the S20, S69, S75, S51 and S82 sample's pickup signal respectively; S9 section deals with mathematical calculation of the pickup voltage; S10 section shows skin depth calculation of $Bi_2Se_3$; S11 section is the image of EM response of $Bi_2Se_3$ for lower and higher frequencies; S12 section deals with position of Fermi level of $Bi_2Se_3$.

47. N. Ni, M. E. Tillman, J.-Q. Yan, A. Kracher, S. T. Hannahs, S. L. Bud'ko and P. C. Canfield, Effects of Co substitution on thermodynamic and transport properties and anisotropic $H_{c2}$ in $Ba(Fe_{1-x}Co_x)_2As_2$ single crystals, Phys. Rev. B **78**, 214515 (2008).

48. B Bag, K Vinod, A Bharathi and S. S. Banerjee, Observation of anomalous admixture of superconducting and magnetic fractions in $BaFe_{2-x}Co_xAs_2$ single crystals, New. J. Phys. **18**, 063025 (2016).

49. Liang Wu, M. Brahlek, R. Valdés Aguilar, A. V. Stier, C. M. Morris, Y. Lubashevsky, L. S. Bilbro, N. Bansal, S. Oh and N. P. Armitage, A sudden collapse in the transport lifetime across the topological phase transition in $(Bi_{1-x}In_x)_2Se_3$, Nat. Physics **9**, 410 (2013).

50. M. Hajlaoui, E. Papalazarou, J. Mauchain, G. Lantz, N. Moisan, D. Boschetto, Z. Jiang, I. Miotkowski, Y. P. Chen, A. Taleb-Ibrahimi, L. Perfetti and M. Marsi, Ultrafast Surface Carrier Dynamics in the Topological Insulator $Bi_2Te_3$, Nano Lett. **12,** 3532 (2012).

51. Dohun Kim, Qiuzi Li, Paul Syers, Nicholas P. Butch, Johnpierre Paglione, S. Das Sarma and Michael S. Fuhrer, Intrinsic Electron-Phonon Resistivity of Bi2Se3 in the Topological Regime, Phys. Rev. Lett. **109**, 166801 (2012).

52. Zhi Ren, A. A. Taskin, Satoshi Sasaki, Kouji Segawa and Yoichi Ando, Large bulk resistivity and surface quantum oscillations in the topological insulator $Bi_2Te_2Se$, Phys. Rev. B **82**, 241306(R) (2010).

53. Wen Jie Wang, Kuang Hong Gao and Zhi Qing Li, Thickness-dependent transport channels in topological insulator $Bi_2Se_3$ thin films grown by magnetron sputtering, Sci. Rep. **6**, 25291 (2016).

54. Matthew Brahlek, Nikesh Koirala, Maryam Salehi, Namrata Bansal, and Seongshik Oh, Emergence of Decoupled Surface Transport Channels in Bulk Insulating $Bi_2Se_3$ Thin Films, Phys. Rev. Lett. **113**, 026801 (2014).

55. G. L. Sun, L. L. Li, X. Y. Qin, D. Li, T. H. Zou, H. X. Xin, B. J. Ren, J. Zhang, Y. Y. Li and X. J. Li, Enhanced thermoelectric performance of nanostructured topological insulator $Bi_2Se3$, Appl. Phys. Lett. **106**, 053102 (2015).
56. Yoichi Ando, Topological Insulator Materials. J. Phys. Soc. Jpn. **82**, 102001 (2013).
57. D. Schoenberg, Magnetic Oscillations in Metals (Cambridge University Press, London) 1984, p. 66.
58. Kazuma Eto, Zhi Ren, A. A. Taskin, Kouji Segawa and Yoichi Ando, Angular-dependent oscillations of the magnetoresistance in $Bi_2Se_3$ due to the three-dimensional bulk Fermi surface, Phys. Rev. B **81**, 195309 (2010).
59. Marco Bianchi, Dandan Guan, Shining Bao, Jianli Mi, Bo Brummerstedt Iversen, Philip D.C. King and Philip Hofmann, Coexistence of the topological state and a two-dimensional electron gas on the surface of $Bi_2Se_3$, Nat. Comm. **1**, 128 (2010).
60. Matthew Brahlek, Nikesh Koirala, Namrata Bansal and Seongshik Oh, Transport Properties of Topological Insulators: Band Bending, Bulk Metal-to-Insulator Transition, and Weak Anti-Localization, Solid State Communications **215-216**, 54–62 (2015).

# A non-contact mutual inductance based measurement of an inhomogeneous topological insulating state in Bi$_2$Se$_3$ single crystal with defects

Amit Jash[1], Kamalika Nath[1], T. R. Devidas[2,3], A. Bharathi[2], S. S. Banerjee[1*]

[1]Department of Physics, Indian Institute of Technology, Kanpur 208016, Uttar Pradesh, India; [2]UGC-DAE Consortium for Scientific Research, Kalpakkam-603104, India; [3]Present address; The Racah Institute of Physics, Hebrew University of Jerusalem, Givat Ram, Jerusalem, 91904, Israel

* Email: satyajit@iitk.ac.in.

## S1: Coil details.

In our two coil mutual setup, both coils are dipole coils. The most important point in coil design is that the winding of the excitation and pickup coils should be as close as possible. The excitation coil has four layers and each layer have 36 turns while pickup coil has four layers having 32 turns. Figure (a) shows the schematic of the excitation coil. The bobbin and stand are made of an insulating materials macor to avoid any unwanted signals developed by eddy currents. The material of the bobbin is non-magnetic. Figure (b) and (c) show the inside of the two coil mutual setup. Figure (d) shows that pickup voltage as a function of applied current in the source coil at 65 kHz. Pickup voltage scales linearly with current in the excitation coil. This measurement is carried out at 140 K of S20 sample. The 1 volt drop across the excitation coil corresponds to 153 mA current in the excitation coil. Note that for our setup there is a linear dependence of the pickup signal with the excitation current in the coils shown below. This allows measurements to be done with higher currents in the coil for better sensitivity without changing any features in the frequency or temperature dependence of the measurements. The linear relationship between pickup voltage amplitude and current in excitation coil shows the measurements with different values of current in the excitation coil can be compared with each other due to the linear relationship.

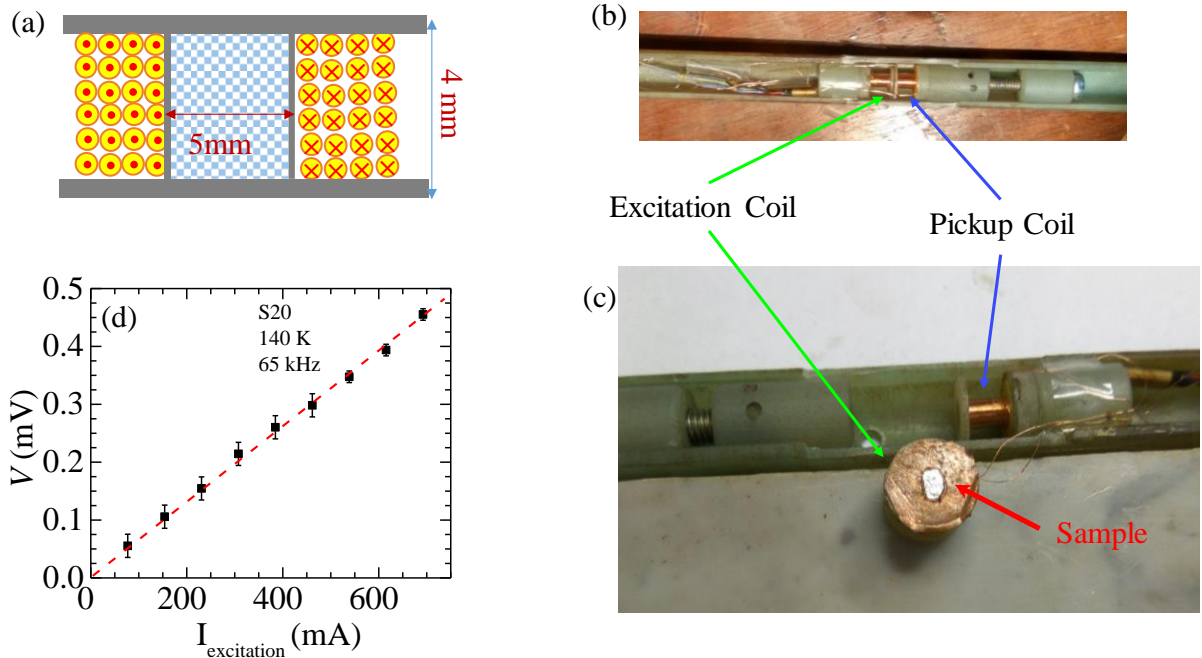

## S2: COMSOL simulation.

We have also verified our experimental data with simulation using a simple model. The simulation part has been performed using Comsol multiphysics software (AC-DC module). The simulation is done by the solving the standard Maxwell EM equations:

$$\left(j\omega\sigma - \omega^2 \varepsilon_0 \varepsilon\right)\vec{A} + \vec{\nabla} \times \frac{\vec{B}}{\mu_0 \mu_r} - \sigma \vec{v} \times \vec{B} = \vec{J}_e$$

$$\vec{\nabla} \times \vec{A} = \vec{B}$$

$$J_e = \frac{N\left(V_{coil} + V_{in}\right)}{R_{coil}}$$

where $\omega$ is the angular frequency of the applied AC signal, $A$ is the magnetic vector potential, $\sigma$ is the conductivity of the material, $v$ is the velocity of the charge particle, $N$ is the number of turns, $R_{coil}$ is the resistance of the coil, $J_e$ is the current density and $V_{coil}$ is the applied AC voltage in the coil. Figure (a) shows the schematic of ideal topological insulator. It also shows dimension of our sample used for the simulation. The sample is 20 μm thick and the two surface states are 5 nm thick each. Figure (b) shows the schematic of inhomogeneous bulk state. The coupling channels are the cylinders having diameter 100 nm. Depending upon the inhomogeneity level, the conducting cylinder concentration is varied keeping the diameter fixed. Note we found no difference in mutual inductance for 30% inhomogeneity (see main text Fig. 6(b)) between an ordered and a disordered configuration of the conducting cylinders. Pickup voltage is sensitive to the fraction of the TI sample being inhomogeneous. Fig. (c) shows inhomogeneous bulk state which has disorder arrangement. For example, in Figs. (b) and (c) the inhomogeneity is 17%, as the number of cylindrical channels are equal in both cases. The overall schematic of the Comsol simulation is shown in Fig. (d).

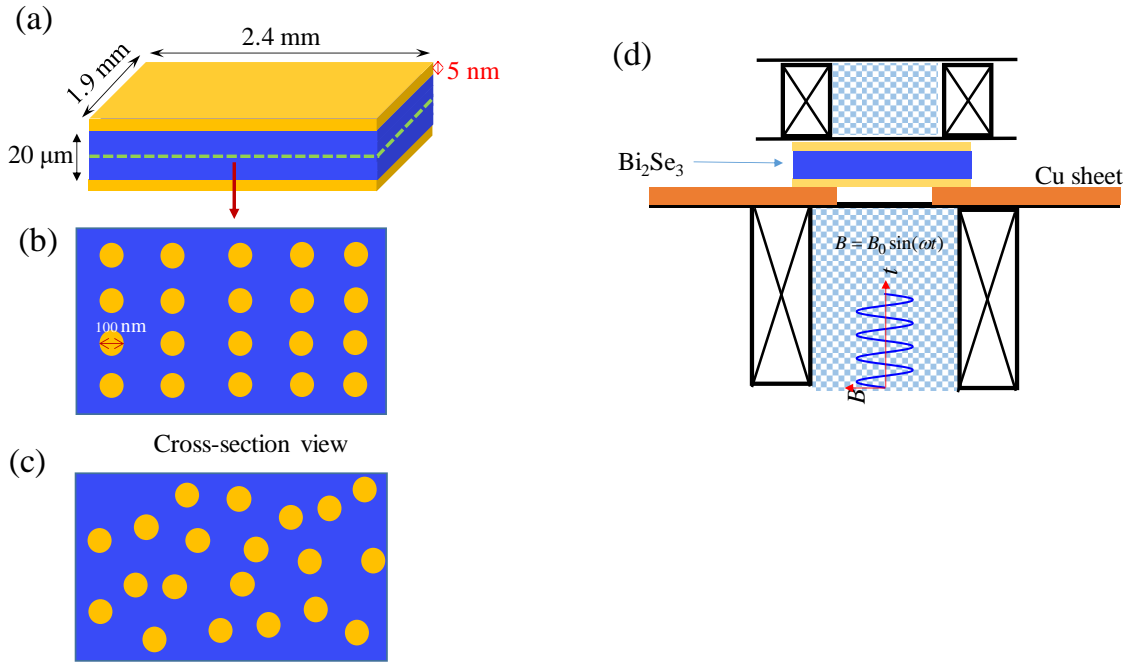

## S3: Background mutual inductance of the two coil setup and homogeneous Cu plate signal.

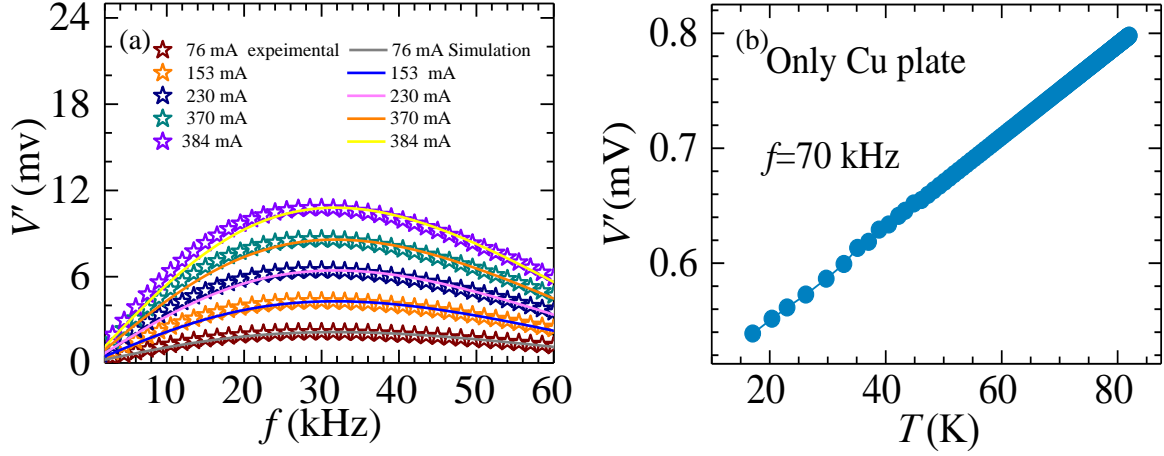

Figure (a) shows the real component of the background signal for different amplitude as a function of frequency. A comparison of the measured signal of only a Cu sheet with the hole placed between the two coils with the signal calculated by solving the Maxwell equation for this complex three dimensional configuration on Comsol. The symbols in figure represent data points of the signal measured by the pickup coil in our two coil setup as a function of frequency at 300 K for different excitation current sent in the primary coil. The solid lines through the data points is the result of calculated signal by solving Maxwell equation for this three dimensional geometry using Comsol. Clearly the close match between the experimental data and the calculations shows the validity of not only the equations we have used for analyzing our problem but also the accuracy of the solutions given by Comsol. Figure (b) displays the pickup voltage as function of temperature when a solid Cu plate (without a hole, thickness 2 mm) is placed between the coil. The pickup voltage from Copper increases linearly with $T$. It can be clearly seen that the behavior of the pickup signal from Cu is completely different from that of the $BaFe_{1.85}Co_{0.15}As_2$ superconducting sample and the $Bi_2Se_3$ TI sample which are discussed in the main paper. In our measurements we have infact used a Copper sheet with a hole in it over which the sample is placed. This would further reduce the amount of Copper within the pickup area of the coil. From here it is clear that the presence of Copper between the coil doesn't affect the signal measured from samples we are investigating.

## S4: S20 sample response.

Figure (a) shows the V(f) for the $Bi_2Se_3$ sample (S20) at 140 K, where $V(f) = \sqrt{V'^2(f) + V''^2(f)} \approx V'(f)$, as for these TI crystals, the $V''(f) \ll V'(f)$. One can identify two regimes of behavior in V(f), namely, one in the low frequency regime where $V(f) \propto f^2$ (see red dotted line through the data), and the other in the higher frequency regime where, $V(f) \propto f^\alpha$; where α close to one (see black solid line to the data). Inset of Fig. (a) shows the V(f) data replotted in a normalized log-log scale, showing the change in slope from quadratic to linear behavior. In the inset, the slope of the red line (quadratic regime) is 1.76 ± 0.05 and black line (linear regime) is 0.88 ± 0.01. To understand the source of change in curvature of V(f), we measure the pickup voltage response at fixed f with varying T. Figure (b) shows the V(T) behavior in S20 sample at two different frequencies of 5 kHz and 65 kHz. The 65 kHz data shows the pickup voltage is nearly constant upto 35 K after which decreases with increasing temperature and at a higher temperature (> 180 K) the V(T) again becomes nearly constant at ~ 0.054 mV. However, at 5 kHz the voltage profile exhibits a completely different behavior compared to 65 kHz data, the V increases exponentially with T and saturates beyond 90 K. The 5 kHz and 65 kHz data merge at a higher temperature. The frequency dependence of V(f) at low T in Fig. (c) is striking, it shows the V(f) ∝ f over the entire frequency range. Figure (d) shows that from 50 K, the nonlinear $V(f) \propto f^2$ dependence gradually develops in the lower f range while at a higher frequency range still a linear frequency dependent regime is maintained. Therefore, below 50 K, V(f) has a predominantly linear dependence on f while at higher T above 50 K V(f) is an admixture of quadratic and linear dependence on f.

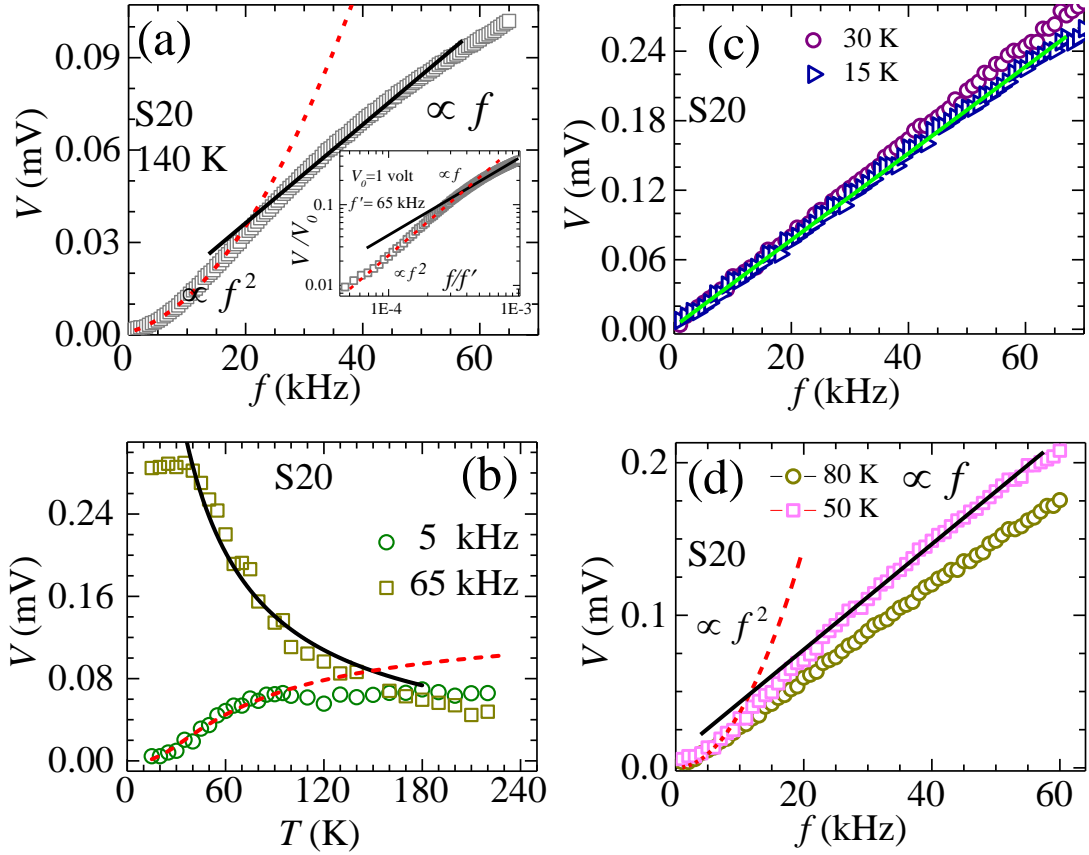

FIG. (a) Variation of pickup voltage ($V$) with frequency for Bi$_2$Se$_3$ single crystal (S20) at 140 K. The inset shows $V/V_0$ vs $f/f'$ in log-log scale. For normalizing the signals, we use $V_0$ = 1 V which is the amplitude of the AC voltage generated across the excitation coil and $f'$ = 65 kHz. All data in the figures correspond to an excitation current of 153 mA in the excitation coil of the setup. Two lines show two distinct regions. Red dotted line is used for the bulk dominating fitting and black line is used for the surface part fitting. We have maintained the same color and style throughout the paper. (b) Variation of pickup voltage ($V$) with temperature ($T$) for S20 at two frequencies 5 kHz and 65 kHz. In 65 kHz data, $V \propto 1/(C'+D'T)$ is fitted where $C'$ = 0.5884 mV$^{-1}$, $D'$ = 0.772 mV$^{-1}$K$^{-1}$ and $V \propto V_{b0} exp(-\Delta/K_bT)$ is fitted in 5 kHz data where $\Delta$ = (5.77 ± 0.45) meV and $v_{b0}$ = 0.15 mV. (c) Frequency variation of $V$ is shown for two temperatures 15 K and 30 K for the same sample. Solid green line is the best fit line through the data points, which shows a linear behavior at low temperatures. (d) $V$-$f$ behavior is shown at higher temperature (50 K & 80 K). Here quadratic and linear fittings are shown with red dash line and black solid line respectively.

## S5: S69 sample response.

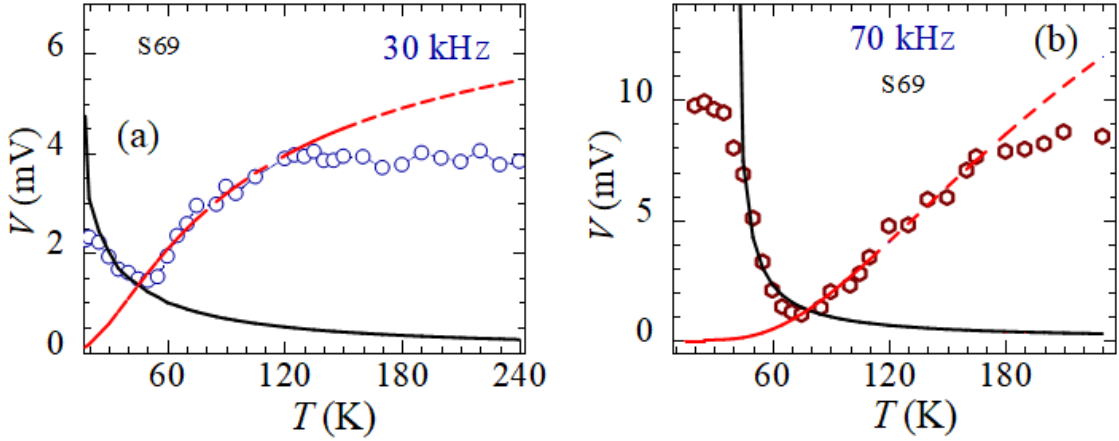

Variation of pickup voltage ($V$) with temperature ($T$) is shown for S69 sample at two frequencies 30 kHz and 70 kHz. The 30 kHz and 70 kHz profiles almost behave similarly with the 65 kHz data (Fig. 5(a) in main manuscript). The red dash line is fitted with, $V \propto V_{b0} \exp\left(-\Delta/K_bT\right)$ and the black solid line fitted with $V \propto 1/(C'+D'T)$. The fitting parameters are shown in the table.

| Frequency | 30 kHz | 70 kHz |
|---|---|---|
| $V_{bo}$ | 7.45 mV | 37.5 mV |
| $\Delta$ | 9.8 meV | 25.6 meV |

| Frequency | 30 kHz | 70 kHz |
|---|---|---|
| $C'$ | 0.0282 mV$^{-1}$ | 0.0062 mV$^{-1}$ |
| $D'$ | 0.082 mV$^{-1}$K$^{-1}$ | 0.199 mV$^{-1}$K$^{-1}$ |

## S6: S75 sample response.

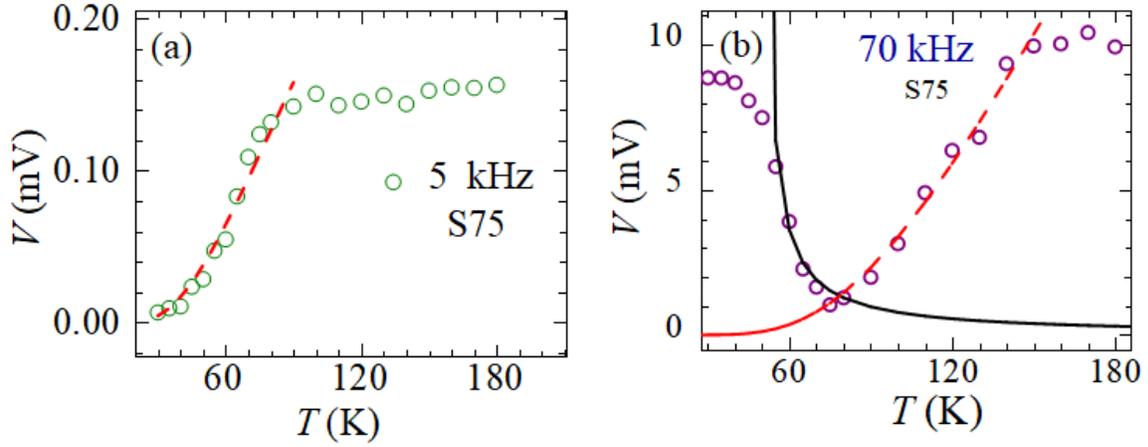

Variation of pickup voltage ($V$) with temperature ($T$) is shown for S75 ample at two frequencies 5 kHz and 70 kHz. The 70 kHz profile is almost similar to the 65 kHz data (Fig. 5(a) in main manuscript). The red dash line is fitted with, $V \propto V_{b0} \exp\left(-\Delta/K_bT\right)$ and the black solid line is fitted with $V \propto 1/(C'+D'T)$. The fitting parameters are shown in the table.

| Frequency | 5 kHz | 70 kHz |
|---|---|---|
| $V_{bo}$ | 0.948 mv | 97.9 mV |
| $\Delta$ | 11.5 meV | 28.6 meV |

| Frequency | 70 kHz |
|---|---|
| $C'$ | 0.00052 mV$^{-1}$ |
| $D'$ | 0.026 mV$^{-1}$K$^{-1}$ |

## S7: S51 sample response.

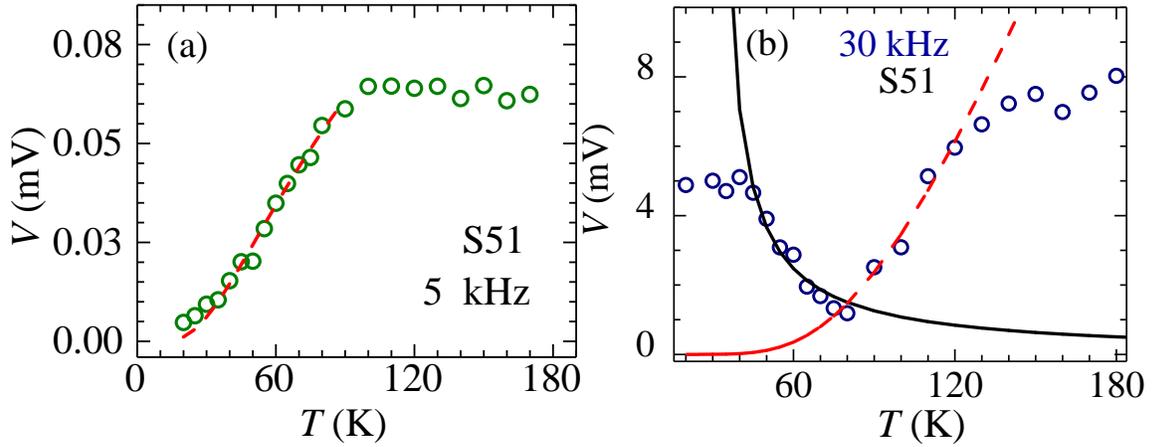

Variation of pickup voltage ($V$) with temperature ($T$) is plotted for S51 sample at two frequencies 5 kHz and 30 kHz. The 30 kHz profile shows almost similar behavior with 65 kHz data (Fig. 5(a) in main manuscript). The red dash line is fitted with, $V \propto V_{b0} \exp\left(-\Delta/K_bT\right)$ and black solid line is fitted with $V \propto 1/(C'+D'T)$. The fitting parameters are shown in the table.

| Frequency | 5 kHz | 30 kHz |
|---|---|---|
| $V_{bo}$ | 0.19 mV | 106.8 mV |
| $\Delta$ | 8.52 meV | 24.12 meV |

| Frequency | 30 kHz |
|---|---|
| $C'$ | 0.00038 mV$^{-1}$ |
| $D'$ | 0.0162 mV$^{-1}$K$^{-1}$ |

## S8: S82 sample response.

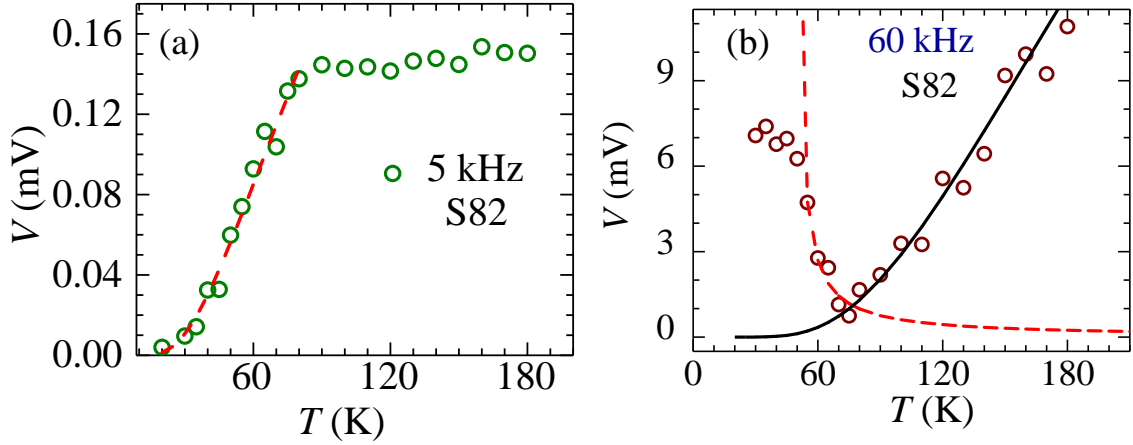

Variation of pickup voltage (V) with temperature (T) is shown for S82 sample at two frequencies 5 kHz and 60 kHz. The 60 kHz profile is almost similar to the 65 kHz data (Fig. 5(a) in main manuscript). The red dash line is fitted with, $V \propto V_{b0} \exp\left(-\Delta/K_b T\right)$ and black solid line is fitted with $V \propto 1/(C'+D'T)$. The fitting parameters are shown in the table.

| Frequency | 5 kHz | 60 kHz |
|---|---|---|
| $V_{bo}$ | 0.675 mV | 121 mV |
| Δ | 10.2 meV | 32.8 meV |

| Frequency | 60 kHz |
|---|---|
| C' | 0.0058 mV$^{-1}$ |
| D' | 0.0016 mV$^{-1}$K$^{-1}$ |

## S9: Pickup voltage generated in the pickup coil.

To explain the pickup data, we use the following model: The AC current in the primary excitation coil produces an AC magnetic field $B = B_0 e^{-i\omega t}$, where $\omega$ is the angular frequency of the AC signal and $B_0$ is the amplitude of the AC magnetic field. For simplicity the sample is assumed to be cylindrical shape with radius $r$. Due to this AC magnetic field, the current density induced in the sample is $\vec{J}(r,z) = \sigma_t \dfrac{i\omega e^{-i\omega t}}{2\pi r} \Phi(z)\hat{\phi}$, where $\sigma_t$ is total conductivity of the sample and $\Phi(z)$ is the magnetic flux passing through the sample.

The magnetic field is generated in the pickup coil from this induced current in the sample. So, the total output voltage in the pickup coil can be written using Faraday's law:

$$V_{pickup}(\omega) = \dfrac{d}{dt}\iint_{r'}\left[\vec{B}(j,r',\theta',z)\cdot d\vec{S}'\right]$$

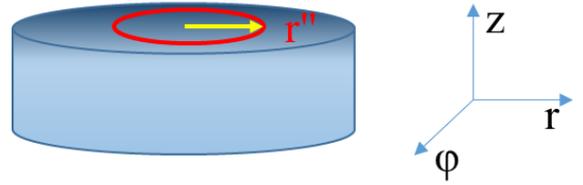

$$= \iint_{r'}\left[\left(\int_{r''}\dfrac{\vec{J}\times\vec{\mathfrak{R}}}{\mathfrak{R}^3}d^3\mathfrak{R}\right)\cdot d\vec{S}'\right] \text{ where } \vec{\mathfrak{R}} = (r'-r'')\hat{r}-\vec{z}$$

$$= -\omega^2 e^{-i\omega t}\sigma_t \iint_{r'}\left[\left(\int_{r''}\dfrac{\Phi(z)\hat{\phi}}{2\pi r''}\times\dfrac{\vec{\mathfrak{R}}}{\mathfrak{R}^3}d^3\mathfrak{R}\right)\cdot d\vec{S}'\right]$$

$$= -\omega^2 e^{-i\omega t}\sigma_t \Phi \xi(r_0,z)$$

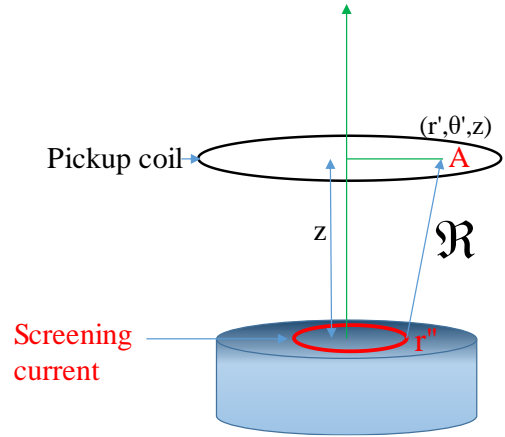

where $\vec{B}(j,r',\theta',z)$ is the magnetic field at the point A in the pickup coil generated by the screening current flowing in the sample. The $\xi(r_0,z)$ is the geometry factor of the mutual coil setup between the sample and pickup voltage

$$V_{pickup}(\omega) = \xi(r,z)\Phi\omega^2 e^{-i\omega t}\sigma_t\dfrac{1}{1+i\omega\tau}$$

Here we have used the Drude like AC conductivity for the surface as well as bulk state of the TI. Herer we have assumed that the distance between the two coils is much larger than the sample thickness. If we consider the temperature dependent conductivity, the total output voltage becomes

$$V_{output}(\omega) = \xi(r,z)\Phi\omega^2 e^{-i\omega t}\left[\dfrac{\sigma_{os}}{C+DT}+\dfrac{\sigma_{ob}}{e^{\Delta/KT}}\right]\dfrac{1}{1+i\omega\tau} \qquad [1]$$

where $\xi(r, z)$ is a geometric factor which is a function of the pickup coil radius ($r$), the height between the two coils ($z$). To understand the frequency and temperature dependence of $V$, we use the parallel transport channel model for TI, where the total electrical conductivity ($\sigma_t$) of the TI is $\sigma_t = \sigma_s + \sigma_b$, where $\sigma_s$ is the surface conductance and $\sigma_b$ is the thermally activated bulk conductance of the bulk. We use the known temperature dependence of the surface ($\sigma_s$) and bulk ($\sigma_b$) conductivities of TI, viz., $\sigma_s = 1/(C+DT)$, where $C$ is related to static disorder scattering and $D$ to electron-phonon coupling strength and $\sigma_b(T) = \sigma_{b0} \exp(-\Delta/K_b T)$, where $\Delta$ is the activation energy scale and $\sigma_{b0}$ is the high temperature conductance of the bulk state. Assuming Drude's AC conductivity for a TI we calculate the frequency dependence of the pickup voltage, $V(f)$ in the low and high frequency regime. In the low frequency limit of the AC excitation ($\omega = 2\pi f$), the alternating magnetic field penetrates the bulk of the sample. Therefore, the induced current ($I_{induced}$) is flowing throughout the sample volume. (see schematic of the magnitude of $I_{induced}$ distribution across sample cross-section in Fig. 4(a) in main manuscript).

It is also known form terahertz experiment that the ultrafast dynamics of the carriers in the surface Dirac cone is delayed with respect to the bulk conduction band [J.A. Sobata, *et al.* PRL **108**,117403, (2012), M. Hajlaoui, *et al.* Nano Lett., **12**, 3532 (2012)]. Surface and bulk state have different scattering time scale which we have used here to explain the linear and quadratic frequency dependence. Motivated by this, we have used different limit for surface and bulk state Hence we are using two different limit for the bulk and surface state. For conduction through the bulk of the TI the $\tau$ (average collision time interval of electron with disorder site) is expected to be small due to large probability of collision, hence at low $\omega$ for small $\tau$ values, the approximate condition $\omega\tau \ll 1$ can be used. It can be shown that with this condition in Eq. **1** one gets, $V(f) \propto f^2$ for low frequencies (namely when response is associated with conduction through the bulk of the TI)

$$V_{pickup}(f) \approx \Phi_0 \xi(r,z) f^2 e^{-i2\pi ft} \sigma_b \qquad [2]$$

In the high frequency ($\omega$) range as the AC field penetrates mainly the surface of the sample, the pickup signal is associated with the response from currents induced close to the surface of the TI. Therefore, induced current ($I_{induced}$) is flowing at the surface only (a schematic of the magnitude of the distribution of $I_{induced}$ across TI sample cross-section is shown in Fig. 4(b) in main manuscript). Therefore for high $\omega$ and other condition $\omega\tau \gg 1$, due to which from Eq. **1** we get $V(f) \propto f$ for the surface state of a TI, viz.,

$$V_{pickup}(f) \approx \Phi_0 \xi(r,z) e^{-i2\pi ft} \sigma_s f \qquad [3]$$

## S10: Estimate of skin depth calculation for Bi$_2$Se$_3$.

Figure shows the variation of skin depth ($\delta$) as function of $f$ at room temperature using the high conductivity of the surface state in Bi$_2$Se$_3$ (viz., $\sigma_s \sim 10^{11}$ S/m). Note that in a TI a high electrically conducting sheath completely covers a much lower electrically conducting medium. An electromagnetic (EM) wave impinging on the TI will always first encounter the high conductivity surface sheath on the TI. Hence the attenuation and consequently the EM skin depth will be determined by the electrical conductivity of the high conducting surface sheet. We show in S11 in supplementary section a simulation showing the attenuation of an impinging EM signal is governed by the high conducting surface sheath in the TI material. In a frequency dependent measurement at moderately low $T$, at high frequencies as the skin depth is already smaller than the sample thickness, in this regime changing the frequency at constant temperature (especially at low $T$), leads to variations in skin depth in the nanometer range. The inset of figure, we plot the behavior of the estimated decrease in skin depth ($\Delta\delta = \frac{d\delta}{df}\Delta f$, for a $\Delta f$ = 500 Hz increase in frequency) versus frequencies. The $\Delta\delta$ is seen to change in the range of a 100 nm and below above 20 kHz. Here were would like to mention that temperature dependent measurement of pickup voltage is not sensitive enough at high $T$ however a frequency dependent measurements at high $T$ of 150 K (see Fig. 3(a)) still observes a change in frequency dependent behavior. This change in curvature of the pickup voltage is seen in a frequency dependent measurement because the thin high conducting surface state regime begins contributing to the signal along with the comparatively lower conducting bulk.

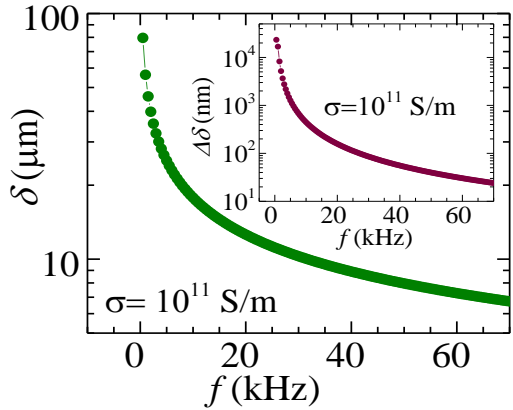

## S11: Bz distribution of the sample.

Figure (a) and (b) are shown the $B_z$ distributions for two different frequency applied in the source coil. The current applied across the source coil is 150mA which is similar to our experiment. In the simulation, we have used the idle topological insulator model without encounter any inhomogeneous fraction in bulk. The schematic of the simulation is shown in Fig. (c) (the cross-section of the sample is shown in Fig. (c), yellow color shows the surface state with conductivity $10^{11}$ S/m and blue color shows the bulk state with conductivity $6\times10^3$ S/m. Bulk state is 20 μm thick and surface state is 10 nm thick). At low frequency, the magnitude of the $B_z$ has much higher value above the sample (Fig. (a)) due to the large skin depth but at a higher frequency the $B_z$ value decreases to ~200 Oe at 50 kHz above the sample surface (Fig. (b)). Now we replace the $Bi_2Se_3$ sample with another normal sample which has same conductivity in bulk and surface ($6\times10^3$ S/m). Figure (d) and (e) are shown the $B_z$ distributions for two different frequency applied in the source coil. From these two figures, we can notice that the $B_z$ distributions are same for these two cases as the skin depths are mm range in these frequency range. From these simulation, we can notice that due to the presence of micron thickness sample, the $B_z$ distribution changes with frequencies in spite of the bulk skin depth in mm range. So, the high conducting surface state which is extended in nm rang, has strong contributions in the $B_z$ distribution.

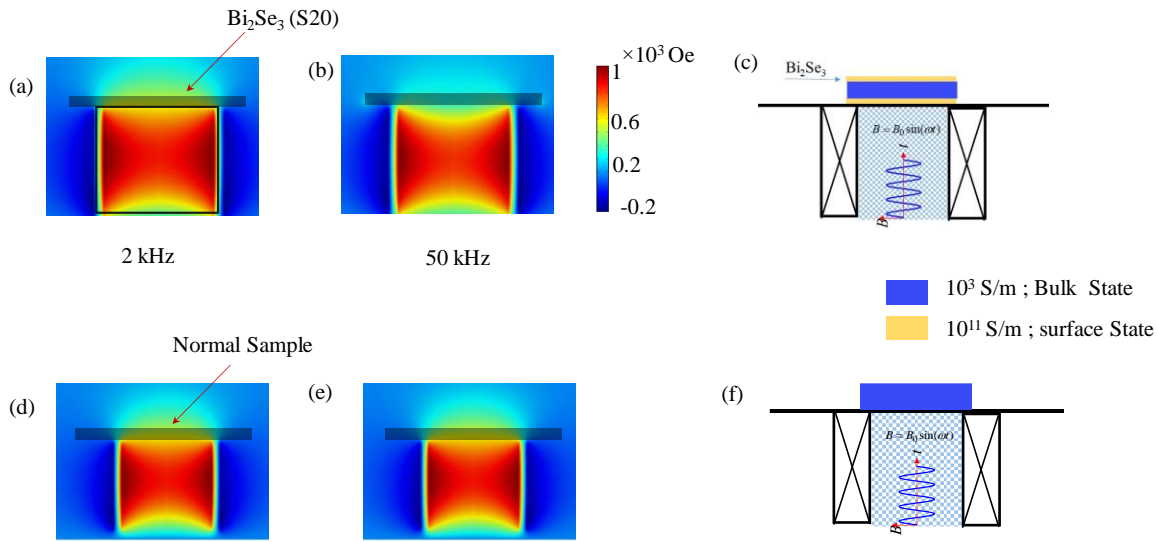

## S12: Position of the Fermi energy.

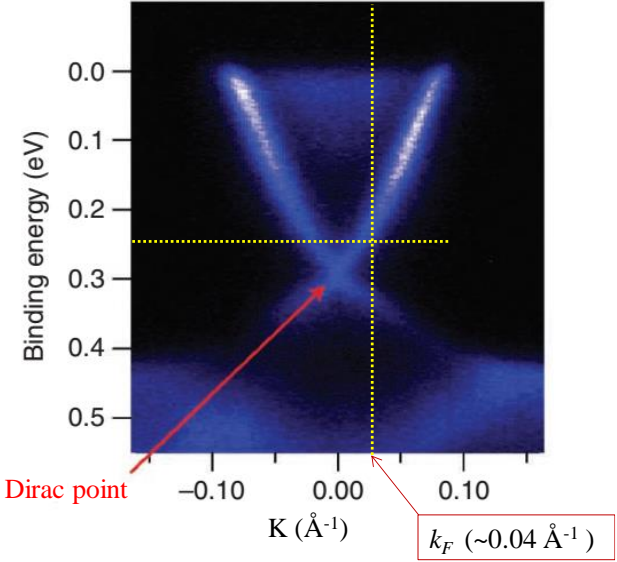

ARPES data reproduced from Fig. 1 of M. Bianchi et al., Nature Comm. 1, 128 (2010)

We use Lifshitz-Kosevich (LK) equation is to analyse the SdH oscillation seen in our $Bi_2Se_3$ sample. From the oscillation period of the SdH oscillations seen in the transport data (Fig. 1a in revised manuscript and see the figure caption), the measured surface carrier density per area is found to be $n_s = (2.268 \pm 0.012) \times 10^{12}$ cm$^{-2}$, which corresponds to a Fermi wave vector for the 2D surface state to be $k_F = \sqrt{2\pi n_s} = (0.0377 \pm 0.0027)$ Å$^{-1}$. By placing the location of $k_F$ on the ARPES spectrum of $Bi_2Se_3$ (see $k_F$ marked by a vertical yellow line in the adjoining figure which is Fig.1 ARPES data of $Bi_2Se_3$ as published by M. Bianchi et al., Nature Comm. 1, 128 (2010)), we see that the Fermi energy is approximately 30 meV above the Dirac point and about 100 meV below the bottom of the bulk conduction band. Such instances of $E_f$ lying within the bulk gap is not unusual and have been reported earlier in $Bi_2Se_3$, for example, J.G. Analytis, R.D. McDonald, S.C. Riggs, J.-H. Chu, G.S. Boebinger, I.R. Fisher, Nat. Phys. 6, 960 (2010); J.G. Analytis, J.-H. Chu, Y. Chen, F. Corredor, R.D. McDonald, Z.X. Shen, I.R. Fisher, Phys. Rev. B: 81, 205407 (2010); M. Brahlek, N. Koirala, M. Salehi, N. Bansal, S. Oh, Phys. Rev. Lett. 113 (2014) 026801, our Ref. 35, T. R. Devidas et al *Euro Phys. Lett.* 108, 67008 (2014). In Topological insulators there, is a significant difference between the surface and bulk Fermi levels. The final location of the Fermi level depends on the charge transfer to make Fermi level equal everywhere. The charge transfers from the bulk to surface or vice-versa creates a potential difference which leads to shifting of bands or band bending effects. An upward or downward band bending depends strongly on the presence of defects in the material and in the present case the upward band bending is related to the presence of Selenium vacancies in the $Bi_2Se_3$ samples as suggested by the study in Ref. 35.